\documentclass[traditabstract]{aa}
\usepackage{graphicx}
\usepackage{txfonts}
\usepackage{longtable}
\usepackage[authoryear]{natbib}
\bibpunct{(}{)}{;}{a}{}{,}
\def\ltsima{$\; \buildrel < \over \sim \;$}
\def\simlt{\lower.5ex\hbox{\ltsima}}
\def\gtsima{$\; \buildrel > \over \sim \;$}
\def\simgt{\lower.5ex\hbox{\gtsima}}

\newcommand{\HI}{\ion{H}{i}}

\begin{document}
   \title{The {\em StEllar Counterparts of COmpact high velocity clouds} (SECCO) survey. II.}
\subtitle{Sensitivity of the survey and an Atlas of Synthetic Dwarf Galaxies.\thanks{Based on data acquired using the Large Binocular Telescope (LBT).
The LBT is an international collaboration among institutions in the United
States, Italy, and Germany. LBT Corporation partners are The University of
Arizona on behalf of the Arizona university system; Istituto Nazionale di
Astrofisica, Italy; LBT Beteiligungsgesellschaft, Germany, representing the
Max-Planck Society, the Astrophysical Institute Potsdam, and Heidelberg
University; The Ohio State University; and The Research Corporation, on behalf
of The University of Notre Dame, University of Minnesota and University of
Virginia.}}

   \author{ G. Beccari\inst{1}, M. Bellazzini\inst{2},  G. Battaglia\inst{3,9}, R. Ibata\inst{4}, N. Martin\inst{4,5}, V. Testa\inst{6},  M. Cignoni\inst{7}, M. Correnti\inst{7}
           }
         
      \offprints{G. Beccari}

   \institute{European Southern Observatory, Karl-Schwarzschild-Strasse 2, D-85748 Garching bei Munchen, Germany
             \email{gbeccari@eso.org} 
             \and
  			INAF - Osservatorio Astronomico di Bologna,
              Via Ranzani 1, 40127 Bologna, Italy
              \and
              Instituto de Astrof'sica de Canarias, 38205 La Laguna, Tenerife, Spain
             \and
             Observatoire astronomique de Strasbourg, Universit\'e de Strasbourg, CNRS, UMR 7550, 11 rue 
             de l'Universit\'e, F-67000 Strasbourg, France 
             \and
             Max-Planck-Institut f\"ur Astronomie, K\"onigstuhl 17, D-69117 Heidelberg, Germany 
             \and
             INAF - Osservatorio Astronomico di Roma, via Frascati 33, 00040 Monteporzio, Italy
             \and
             Space Telescope Science Institute, 3700 San Martin Drive, Baltimore, MD 21218, USA
             \and 
             INAF - Osservatorio Astrofisico di Arcetri, Largo E. Fermi 5, I-50125, Firenze, Italy
              \and 
			Universidad de la Laguna, Dpto. Astrofisica, E-38206 La Laguna, Tenerife, Spain
          }

     \authorrunning{G. Beccari et al.}
   \titlerunning{The SECCO Survey. II. Sensitivity and an Atlas of Synthetic Dwarf Galaxies.}

   \date{Submitted to A\&A }

\abstract{SECCO is a survey devoted to the search for stellar counterparts within Ultra Compact High Velocity Clouds that are candidate low mass / low luminosity galaxies. In this contribution
we present the results of a set of simulations aimed at the quantitative estimate of the sensitivity of the survey as a function of the total luminosity, size and distance of the stellar systems we are looking for. For all our synthetic galaxies we assumed an exponential surface brightness profile and an old and metal-poor population. 
The synthetic galaxies are simulated both on the images and on the photometric catalogs, taking into account all the observational effects. In the fields where the available observational material is of the top quality ($\simeq$ 36\% of the SECCO fields) we detect synthetic galaxies as $\ge 5\sigma$ over-densities of resolved stars down to $\mu_{V,h}\simeq 30.0$~mag/arcsec$^2$, for D$\le 1.5$~Mpc, and down to $\mu_{V,h}\simeq 29.5$~mag/arcsec$^2$, for D$\le 2.5$~Mpc. In the field with the worst observational material of the whole survey we detect synthetic galaxies with $\mu_{V,h}\le 28.8$ ~mag/arcsec$^2$ out to D$\le 1.0$~Mpc, and those with $\mu_{V,h}\le 27.5$ ~mag/arcsec$^2$ out to D$\le 2.5$~Mpc. 
Dwarf galaxies with $M_V=-10.0$, with sizes in the range spanned by known dwarfs, are detected by visual inspection of the images up to D=5~Mpc independently of the image quality. In the best quality images dwarfs are partially resolved into stars up to D=3.0~Mpc, and completely unresolved at D=5~Mpc. 
As an independent test of the sensitivity of our images to low surface brightness galaxies we report on the detection of several dwarf spheroidal galaxies probably located in the Virgo cluster with $M_V\la -8.0$ and $\mu_{V,h}\la 26.8$~mag/arcsec$^2$. The nature of the previously discovered SECCO~1 stellar system, also likely located in the Virgo cluster, is re-discussed in comparison with these dwarfs. While specific for the SECCO survey, our study may also provide general guidelines for detection of faint stellar systems with 8m class telescopes.}


\maketitle
%
%________________________________________________________________

\section{Introduction}
\label{intro}

Within the standard $\Lambda$-Cold Dark Matter (CDM) cosmology it is generally accepted that baryonic physics has a significant impact in shaping the evolution of dwarf galaxies. All the relevant processes (e.g., re-ionization, Supernova feedback, ram-pressure stripping) act by preventing/quenching star formation in low-mass halos \citep[$M\le 10^9~M_{\odot}$, mini-halos, after][and references therein]{ricotti}. Within this framework, the inclusion of baryonic physics in CDM N-body models of galaxy formation can strongly mitigate the long standing {\em missing satellites} problem \citep[][and references therein]{kauff,klip,moore,kra04}. Independently of the actual ability of these models to capture all the complex events leading to galaxy formation, 
their general prediction is that there should be a significant number of dark matter mini-halos containing $10^5-10^7~M_{\odot}$ of baryons in the form of neutral hydrogen, with or without an associated (small) stellar component \citep{ricotti,sawa}. The observational detection of these dark and {\em almost-dark} galaxies 
is a crucial frontier for the validation of the current cosmological  paradigm on the small-scales \citep[see, e.g.,][]{can15}. \\

While modern all-sky surveys allowed us to unveil a large number of very faint gas-deficient galaxies around the Milky Way \citep[the so called Ultra Faint Dwarfs - UFD][]{belo} the hypothesised gas-rich and star-poor dwarfs are still to be found.  All UFDs were found as stellar over-densities; only one of them was subsequently found to have a neutral hydrogen component and ongoing star formation, Leo~T \citep{leoT}. On the other hand the recent identification of the faint star-forming dwarf galaxy Leo~P \citep{leop_1} opened a new path for the discovery of
these systems. In fact, Leo~P was discovered as the stellar counterpart of an Ultra Compact High Velocity Cloud (UCHVC) previously identified by the ALFALFA HI survey \citep{giova07}.
Along this line, \citet[][A13 hereafter]{adams} selected from the ALFALFA database a sample of 59 UCHVCs lacking any visible counterpart in SDSS images, that they proposed as good candidates to be associated with mini-halos in the distance range 0.25~Mpc $\le D \le 2.0$ Mpc. 
Similarly, \citet[][S12 hereafter]{GALFA}, identified 27 UCHVCs from their GALFA-HI survey as local ``galaxy candidates'' \citep[see][for a critical comparison of the two sources of candidates]{sand15}. 
The only way to directly confirm these candidates as genuine dwarf galaxies would be to identify a concomitant stellar counterpart whose distance could be estimated, as done in the case of Leo~P \citep{leop_lbt}.\\  

The search for stellar counterparts in these UCHVCs is the main scientific goal of the SECCO {\em (searching for StEllar Counterparts of COmpact high velocity clouds)} survey\footnote{\tt http://www.bo.astro.it/secco/}, that we presented, together with the results from its first phase, in \citet[][Pap~I hereafter]{secco_p1}. In particular we used the LBC cameras \citep{lbc}, mounted at the Large Binocular Telescope\footnote{\tt http://www.lbto.org} (Mt~Graham, AZ) to obtain deep homogeneous wide-field imaging and photometry in two broad bands (g, r) of 25 UCHVCs carefully selected from the A13 sample, plus Leo~P, taken as a template. In Pap~I we concluded that no obvious stellar counterpart in the range of distances indicated by A13 can be associated to the surveyed UCHVCs. On the other hand we identified two candidates that are expected to lie at larger distance. The follow-up of one of them revealed that it is indeed a very faint and dark star-forming galaxy likely belonging to the Virgo cluster of galaxies \citep[SECCO~1,][]{secco_l1}; this result was independently confirmed by \citet{sand15}.

Several teams worldwide are performing similar surveys \citep[see, e.g.][]{tol15,sand15,donma}. In this context SECCO is the first project (and, the only one, so far) to have performed a systematic homogeneous search, since all the targets have been observed with the same set up, albeit not always with the same observing conditions (see Pap~I for details and discussion). It is also based on the deepest images; our photometry typically reaches $r\simeq 26.5$. Finally SECCO has been built to allow a quantitative estimate of the significance of non-detections. In Pap~I we presented a first limited set of simulations with synthetic galaxies, taking into account all the observational effects, that lead us to conclude that we cannot have missed any dwarf having integrated absolute magnitude $M_V\le -8.0$ and half-light radii
$R_h\le 300$~pc lying within 1.5~Mpc from us. In the present contribution we extend our set of experiments with synthetic galaxies in order to explore the sensitivity limits of our survey in terms of total luminosity, size and distance of the stellar systems we are looking for. The full range of observational conditions encountered in SECCO is also probed. This is an essential complement to SECCO observations, providing the final complete result of our screening, at least for the local volume. Moreover, the results of our experiments, performed both on star counts and directly on images, provide a useful set of cases that can be used as a guideline for interpreting the results from similar surveys and to plan new ones.

%------------------------FIG 1-----------------------------------
   \begin{figure}
   \centering
   \includegraphics[width=\columnwidth]{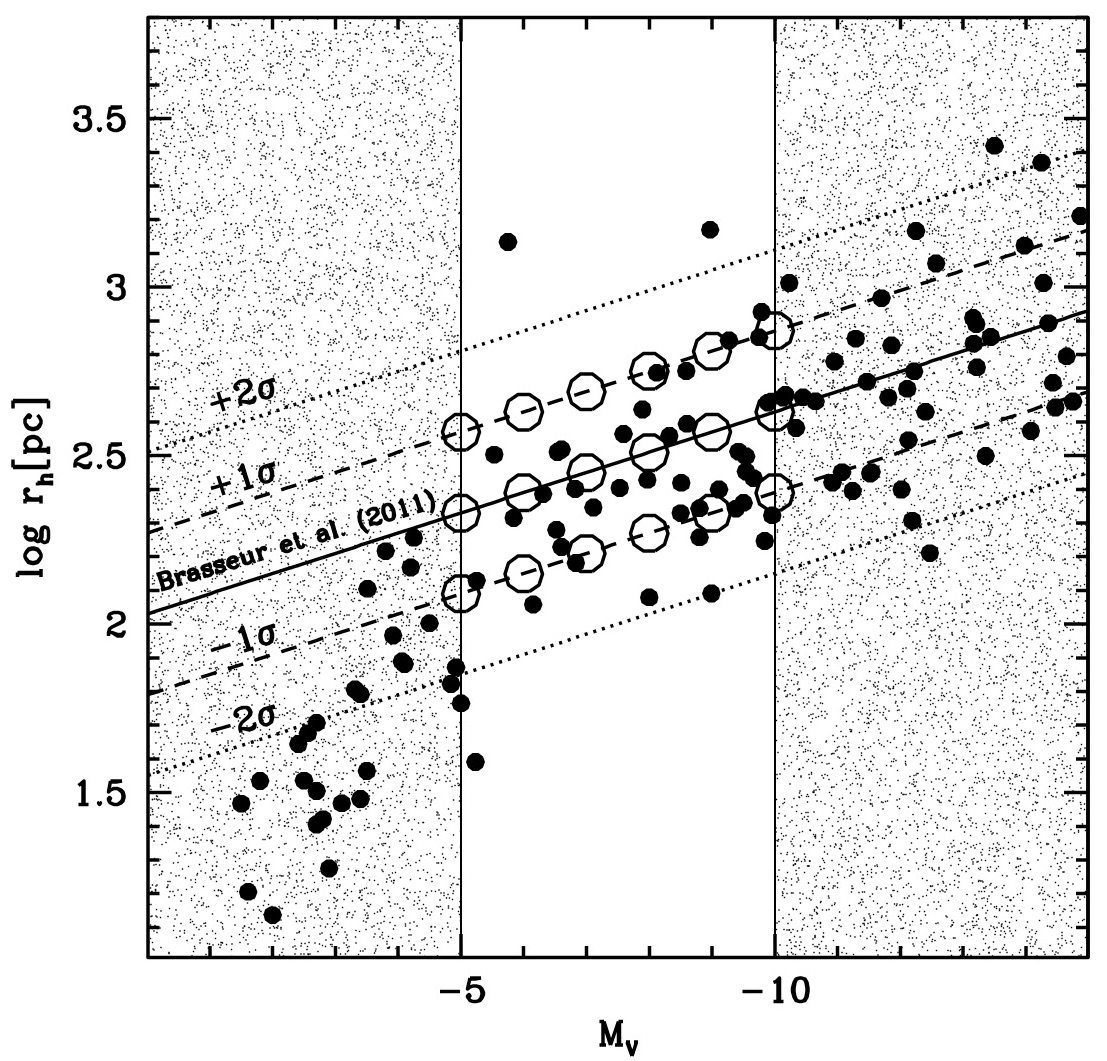}
     \caption{Integrated absolute V magnitude versus the logarithm of the half light radius for dwarf galaxies
     in the most recent version of the M12 catalog (filled circles). The mean $M_V$~vs.~log$R_h$ relation by \citet{brasseur} is plotted as a continuous line;
     the curves bracketing $\pm 1\sigma$ and $\pm 2\sigma$ range about that relation are plotted as dashed and dotted   lines respectively.
     The central not-shaded area of the figure highlights the range of $M_V$ considered in the present study. Large empty circles are the synthetic dwarf galaxies studied in this paper (see also Fig.~\ref{syntpar}).}
        \label{brass}
    \end{figure}

%%%%%%%%%%%%%%%%%%%%%%%%%%%%%%%%%%%%%%%%%%%%%%%%%%%%%%%%%%%%%%%%%

%------------------------FIG 2-----------------------------------
   \begin{figure}
   \centering
   \includegraphics[width=\columnwidth]{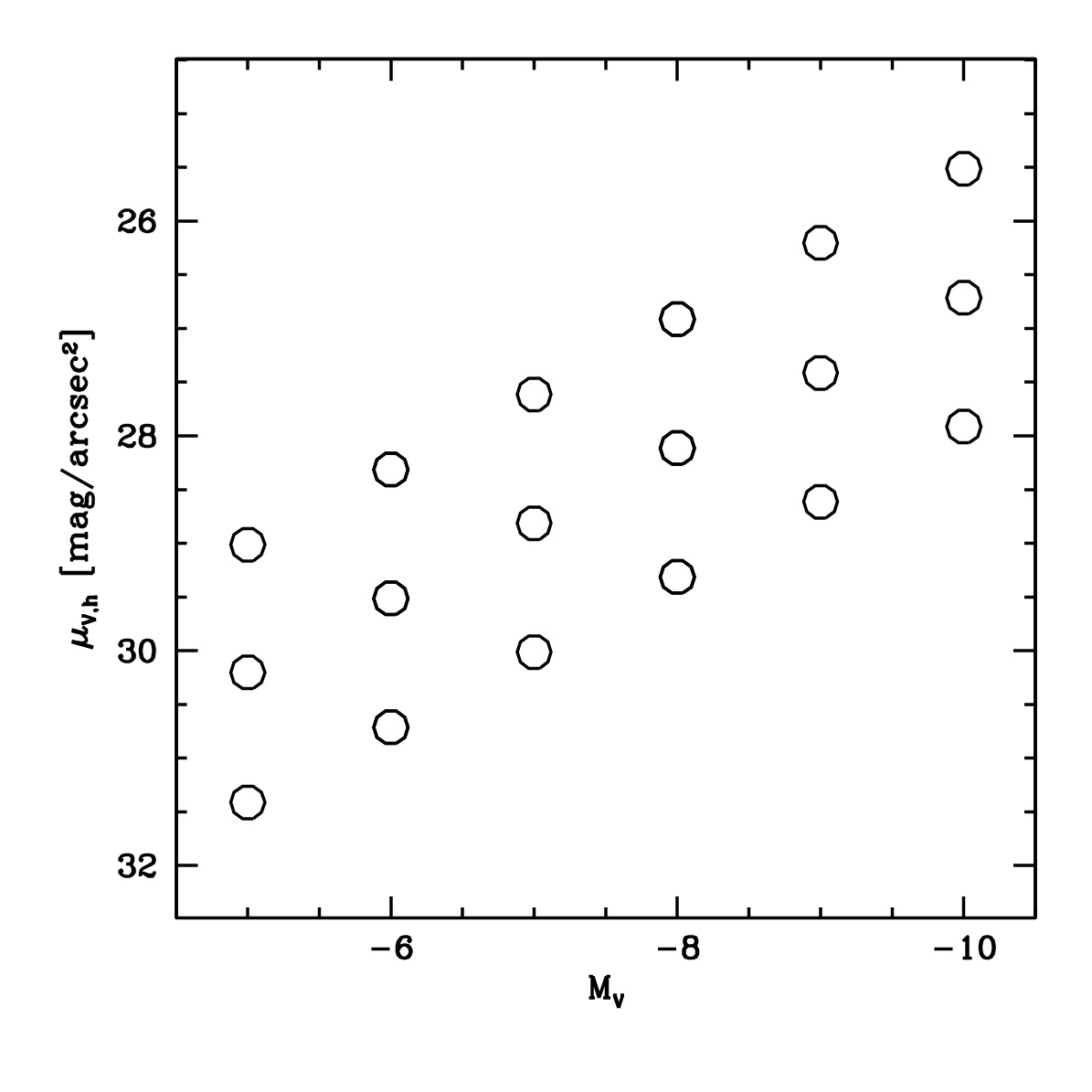}
     \caption{Integrated absolute V magnitude vs. average surface brightness within $R_h$ (upper panel) for the synthetic dwarf galaxies studied in this paper (see also Fig.~\ref{brass}). All the considered dwarfs have an exponential surface brightness profile. }
        \label{syntpar}
    \end{figure}

%%%%%%%%%%%%%%%%%%%%%%%%%%%%%%%%%%%%%%%%%%%%%%%%%%%%%%%%%%%%%%%%%

%------------------------FIG 1-----------------------------------
%   \begin{figure}
%   \centering
%   \includegraphics[width=\columnwidth]{disrad.jpg}
%     \caption{Angular half-light radius ($R_h$) as a function of distance for two assumptions on the physical half-%light radius ($r_h$). A value of $r_h=1.1$~kpc corresponds to the upper limit of the distribution for dwarf galaxies in the Local Volume \citep{mcc} in the range of luminosities that is relevant for the present survey ($M_V\ge -12.0$; with the only exception of Andromeda~XIX, whose structure is likely strongly disturbed by the tidal interaction with M~31, see \citealt{XIX}). There are only nine dwarfs with $r_h\ge 0.5$~kpc among the fifty-four having $M_V>-12.0$; Leo~P has $r_h\sim 0.2$~kpc (N. Martin, private communication).}
%        \label{disrad}
%    \end{figure}

%%%%%%%%%%%%%%%%%%%%%%%%%%%%%%%%%%%%%%%%%%%%%%%%%%%%%%%%%%%%%%%%%

The paper is organised as follows: in Sect.~\ref{syntgal} we describe the synthetic galaxies we adopt in our study. Sect.~\ref{sensmap} explores the sensitivity of the search for stellar over-densities by means of density maps performed in Pap~I. In Sect.~\ref{sensima} we show and discuss our main results on the sensitivity of the visual inspection of images, also performed in Pap~I. In Sect.~\ref{lsb} we report on the discovery of seven Low Surface Brightness dwarfs likely belonging to the Virgo cluster of galaxy and we use these detection to get an independent validation of the sensitivity of our search by visual inspection. Finally, in Sect.~\ref{conc} we summarise the main conclusion of our analysis, also briefly re-discussing the results of Pap-I at the light of the present study. A long table summarising the results of the analysis on density maps is provided in
Appendix~\ref{app_dens}, while the images of synthetic galaxies are collected in 
Appendix~\ref{app_ima}.

%%%%%%%%%%%%%%%%%%%%%%%%%%
%
\begin{table}
  \begin{center}
  \caption{Properties of the synthetic galaxies}
  \label{lista}
  \begin{tabular}{ccc}
$M_V$ &  $R_h$ & $\mu_{V,h}$\\
      &  [pc]  & [mag/arcsec$^2$]  \\
\hline
-10.0 &  245 & 25.5 \\
-10.0 &  426 & 26.7 \\
-10.0 &  741 & 27.9 \\
 -9.0 &  213 & 26.2 \\  
 -9.0 &  371 & 27.4 \\ 
 -9.0 &  645 & 28.6 \\ 
 -8.0 &  186 & 26.9 \\
 -8.0 &  323 & 28.1 \\
 -8.0 &  562 & 29.3 \\
 -7.0 &  162 & 27.6 \\
 -7.0 &  281 & 28.8 \\
 -7.0 &  489 & 30.0 \\
 -6.0 &  141 & 28.3 \\
 -6.0 &  245 & 29.5 \\
 -6.0 &  426 & 30.7 \\
 -5.0 &  123 & 29.0 \\
 -5.0 &  213 & 30.2 \\
 -5.0 &  371 & 31.4 \\
\hline
\end{tabular} 
\tablefoot{All the synthetic galaxies have exponential surface density profiles.
$M_V$ is the integrated absolute magnitude in V band. 
$\mu_{V,h}$ is the average surface brightness within $R_h$.}
\end{center}
\end{table}
%%%%%%%%%%%%%%%%%%%%%%%%%%%%%%%%%%%%%%%%%%%%%%%%%%%%%%%%%%%%%%%%%%%%%%%%%%%%%%%%%%%%%%%%%%%%%%%%%%%%%%%%%%%%

\section{Synthetic galaxies}
\label{syntgal}

The synthetic galaxies that we inject into our photometric catalogs (to test the sensitivity of the detection of  over-densities in stellar density maps) or in our stacked images (to test the sensitivity of the detection by visual inspection) have the same properties as those adopted and described in Pap~I. Here we summarise their main features for the reader's convenience and we illustrate the space of parameters that is explored in the present analysis.

The galaxies consist of a synthetic stellar population of old and metal-poor stars, obtained from the \citet{bressan} models. The mass of the stars are extracted (down to the hydrogen burning limit) from a mass function $N(m)\propto m^{-x}$, with $x=-2.35$ for $m>0.5~M_{\sun}$ and $x=-1.30$ for $m\le 0.5~M_{\sun}$. The stars span a small range in metallicity ($\sigma_{[Fe/H]}=0.1$~dex) with  mean metallicity $[Fe/H]=-1.8$ (see Pap~I).
Star formation starts 13.0~Gyr ago and declines exponentially with time with an exponential scale of
0.5~Gyr. The population is intended to be a fair representation of very faint dwarfs ($M_V\ge -10.0$) that are known to be dominated by old and metal-poor populations \citep[see, e.g.,][]{mateo,mcc,we14}. The choice of a mainly old population is conservative, since the lack of bright blue stars makes its detection more difficult than for a star-forming galaxy. On the other hand the inclusion of a young stellar component in our synthetic galaxies would have implied a strong  dependency on further assumptions, e.g. on star formation efficiency and initial mass function.
%
%
%
%The choice of a mainly old population is conservative. It is a general property of dwarf galaxies that younger stars are more centrally concentrated than old stars \citep{harb,lian}, hence the presence of a young population in a real galaxy would make its detection easier both with density map, enhancing the density at the center with respect to a purely old population, and by visual inspection, since even small groupings of relatively bright blue stars are easier to spot out than a diffuse halo of old stars. In this sense, the detection of SECCO~1 provides an excellent (and extreme) example: the system was independently found by us \citep{secco_l1} and by \citet{sand15} as a tiny group of blue stars without any detectable diffuse old component (see Sect.~\ref{lsb}, for further discussion). 

The synthetic stars are spatially distributed in the plane of the sky according to an exponential profile\footnote{Here we adopt the formalism introduced by \citet{ciotti} for Sersic models, with $R_h=R_e$. The exponential profile is the case $m=1$.}. This is known to be an appropriate choice for a wide range of low-luminosity dwarfs \citep{mateo, martin} and it is very simple, being specified only by the value of its half-light radius $R_h$ and a luminosity normalisation (here the integrated magnitude, $M_V$, that is fixed by extracting a fraction of the synthetic stellar population that sums up the required total luminosity). Circular (spherical) symmetry is always adopted.

We consider only dwarf galaxies fainter than $M_V=-10.0$. The performed experiments reveal that this choice is adequate, since galaxies with $M_V\le-10.0$ are detected in SECCO over the whole range of parameters relevant for our purposes, independently of the quality of the observational material (see Sect.~\ref{sensima}, below). 
On the other hand, our results suggest that dwarfs fainter than $M_V=-5.0$ would be detected only if they are more compact than average and have $D\la 0.5$~Mpc.
The half-light radius of dwarf galaxies is known to scale with their absolute integrated magnitude \citep[see, e.g.,][and references therein]{tht,mcc}.  While this relation is likely significantly shaped by incompleteness, especially on the low surface brightness side and for $M_V\ga -6.0$ \citep{brasseur}, it provides us with a basic guideline for the main goal of the present study, that is to explore the sensitivity of SECCO in the range of {\em known dwarf galaxies}. In Sect.~\ref{sensmap} and Sect.~\ref{conc} we will show that the adopted approach allow us also to probe the sensitivity of the survey also in regions of the parameter space that are still poorly explored by existing panoramic surveys, in particular for $\mu_{V,h}>26.0$~mag/arcsec$^2$ at D$\ga 1.0$~Mpc.

\citet{brasseur} studied this relation in detail and provided a simple model for the mean half-light radius $\bar R_h$ as a function of $M_V$ based on Milky Way and M31 satellites, 
with well measured standard deviations $\sigma_{{\rm log}R_h}$ about the mean relation, taking into account the effect of incompleteness

\begin{equation}
{\rm log}({\bar R_h(M_V)})= 2.39 +0.24(M_V+6.0)   ~~{\rm with}~~\sigma_{{\rm log}R_h}=0.06
\end{equation} 

\citep[see][for details]{brasseur}.
Here, for each value of $M_V$, we consider synthetic galaxies having 
${\rm log}(R_h)={\rm log}({\bar R_h(M_V)})$, 
$\bar R_h(M_V) -1\sigma_{{\rm log}R_h}$, and $\bar R_h(M_v) +1\sigma_{{\rm log}R_h}$, to explore the range of sizes spanned by known dwarfs. 
In the following we will often refer to models with $R_h=\bar R_h(M_v)$, $\bar R_h(M_v) -1\sigma_{{\rm log}R_h}$, and $\bar R_h(M_v) +1\sigma_{{\rm log}R_h}$ as {\em average}, {\em compact}, and {\em diffuse} models, respectively.

In Fig.~\ref{brass} the \citet{brasseur} model (together with its $\pm 1\sigma_{{\rm log}R_h}$ and  $\pm 2\sigma_{{\rm log}R_h}$ envelopes) is compared to the dwarfs in the Local Volume (LV) from the compilation by \citet{mcc}. Note that, in the range of $M_V$ considered here, systems with $R_h\ge \bar R_h(M_v) +1\sigma_{{\rm log}R_h}$ are very rare, while several dwarfs have $R_h\le \bar R_h(M_v) -1\sigma_{{\rm log}R_h}$. This may be partly due to selection effects; however the asymmetry remains also for $M_V\le -8.0$, suggesting that dwarfs more compact than average are more frequent than dwarfs more extended than average.

The grid of $M_V$, $R_h$ and $\mu_{V,h}$\footnote{Average surface brightness within $R_h$ of the corresponding exponential model, derived as $\mu_{V,h} = M_V +5{\rm log}R_h +23.567$, where $R_h$ is expressed in parsecs.} of the synthetic galaxies considered here is reported in Table~\ref{lista} and illustrated in Fig.~\ref{brass} and Fig.~\ref{syntpar}. Knots of the grid have been actually explored only if required. For example, if for a given $M_V$ the dwarfs with $R_h=\bar R_h(M_v)$ are not detected in a given dataset, the exploration of the corresponding  $\bar R_h(M_v) +1\sigma_{{\rm log}R_h}$ case is pointless, since the detection of more diffuse systems is more difficult.

We adopted a different grid of distances for the density maps and visual inspection experiments. The first ones are sensitive only to galaxies that are at least partially resolved into stars, hence relatively nearby, but can reveal also very faint and diffuse systems. For this set of experiments we simulate the synthetic dwarfs at the distance of 0.25, 0.5, 1.0, 1.5, 2.0, and 2.5~Mpc. On the other hand, the visual inspection can reveal also more distant and unresolved systems. In this case we consider the cases D= 0.5, 1.0, 2.0, 3.0, and 5.0~Mpc. A general conclusion from all our experiments with images of synthetic galaxies is that, in agreement with the rule-of-thumb statement made in Pap~I, {\em the distance limit at which we partially resolve dwarfs into stars also in the best observational conditions is $D\simeq 3.0$~Mpc} (see below).

Finally we performed our experiment using (a) a dataset that is representative of the 36 per cent of the SECCO fields observed under the {\em best observing conditions} (Field~B), and (b) the dataset of the field observed under the {\em worst observing conditions} of the whole survey (Field I, see Pap~I). This is intended to bracket the whole range of observing conditions encountered in SECCO and to fully assess the impact of the data quality on the sensitivity of the survey. For reference, we remind that the images of Field B(I) have been obtained with typical seeing of $0.9\arcsec (1.4\arcsec)$, with a limiting magnitude of the final photometry (as traced by the $r_{90}$ parameter, see Pap~I) of $r_{90}=26.37$($r_{90}=25.59$). In the following we will often refer to these two cases as to {\em Best} and {\em Worst} cases, respectively, for brevity and simplicity.

\subsection{Injecting synthetic galaxies into photometric catalogs}
\label{injmap}

The process is described in detail in Pap~I. To summarise we use the extensive sets of
artificial stars experiments presented in Pap~I to transform the catalog of synthetic
stars for a given $M_V$ according to all the effects associated to the observation and data reduction,
i.e. incompleteness, photometric errors, selection effects. 
%In practice we associate
%to each synthetic star (moved to the adopted distance and corrected for the extinction of the %considered
%field) a set of artificial ``sisters'' having similar apparent magnitude and position in the field.
%Then we extract one of these artificial sisters at random and we associated it to the synthetic star %that, consequently, is lost or selected out if this was the case of the artificial star, or, if %survived, has its magnitudes changed according to the changes suffered by the artificial stars in %the process of observation + data reduction.

The synthetic catalog produced in this way is then merged to the original photometric catalog of the fields B and I of the SECCO survey and the surface density maps are derived exactly in the same way as in Pap~I. Then we annotate if the synthetic dwarf is detected at $\ge 5\sigma$ or at $\ge 10\sigma$ above the background or not detected. We also verify if the nature of detected over-densities can be firmly established by the inspection of the Color Magnitude Diagram (CMD) within a given radius (taken as $R_h$, for convenience), compared to the CMD of the surrounding field. This classification process will be illustrated in detail in Sect.~\ref{sensmap}.

\subsection{Injecting synthetic galaxies into stacked images}
\label{injima}

The magnitudes and positions of the stars in the synthetic catalogues described in the previous sections 
are used to inject the synthetic galaxies in the real images with the DAOPHOT/ADDSTAR routine~\citep[][]{daophot}. 
In short, the routine  adds simulated stars into the real images in accordance with their positions and magnitudes, using the Point Spread Function (PSF) 
model characteristic of each image and used for the photometric reduction (see Pap I). 

We have a total of 4 science frames for each field, namely 2 $g$ and 2 $r$ band exposures. The simulated stars are first added to the best seeing
exposure taken as the reference frame. The center of the synthetic  object ( $X_0$,$Y_0$) is placed roughly at the center of the reference frame
and far from saturated stars in order avoid contamination from scattered light\footnote{Note that heavily saturated stars would prevent the detection of a background dwarf if the size of the circle dominated by their light on the image is comparable to the size of the dwarf. The synthetic galaxies considered here typically have $r_h\ga10\arcsec$ (up to $>60\arcsec$). Identifying bright stars in our images we found that stars with r$\la 12.0$ are required to significantly reduce the sensitivity over a circle with radius $\ga 15\arcsec$; stars with r$\la 13.0$ obliterate circles with radius $\ga 10\arcsec$, those with with r$\la 15.0$ obliterate circles with radius $\ga 5\arcsec$. These bright stars are rare in our high latitude fields (19 of 25 having galactic latitude $>50\degr$). 
Using the TRILEGAL galactic model \citep{trilegal} we find that the density of r$\la 13.0$ stars in a field with (l,b)=(90$\degr$,45$\degr$) is $\simeq$ 1 per 40 arcmin$^2$. Hence the impact of chance superposition between a dwarf galaxy and such bright stars should not have a serious impact on our estimate of the sensitivity.}. 
The physical position of each star in the image is simply calculated as $X_s=X_0+X_c$ and $Y_s=Y_0+Y_c$, 
where $X_s$ and $Y_s$ are the coordinates with respect to the coordinates in pixel of the center $X_c,Y_c$.

The 4 images with the synthetic stars were then fed to MONTAGE2~\citep[][]{daophot} in order to build a stacked image
for each simulation. The latter are used to  assess the sensitivity of the SECCO survey through the visual inspections (see Sect.~\ref{sensima}).

\section{Sensitivity of the density maps}
\label{sensmap}

%------------------------FIG -----------------------------------
   \begin{figure}
   \centering
   \includegraphics[width=\columnwidth]{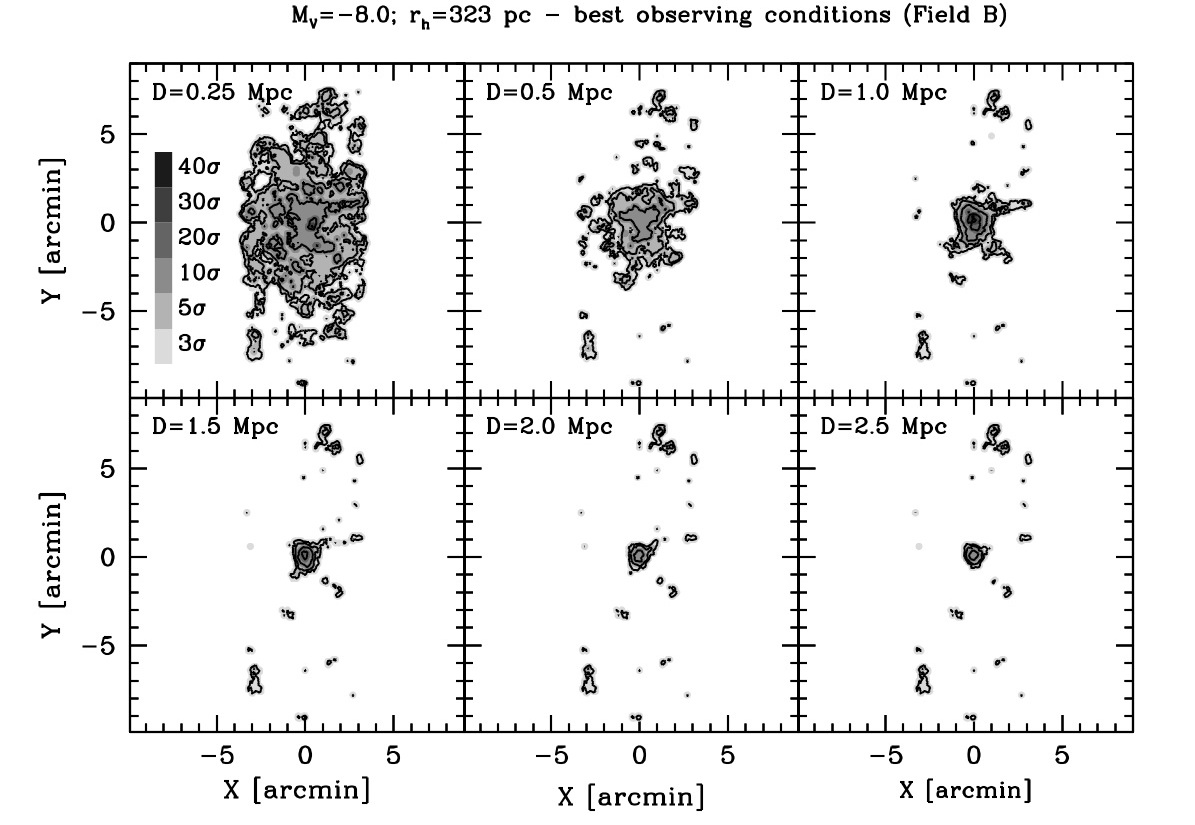}
   \includegraphics[width=\columnwidth]{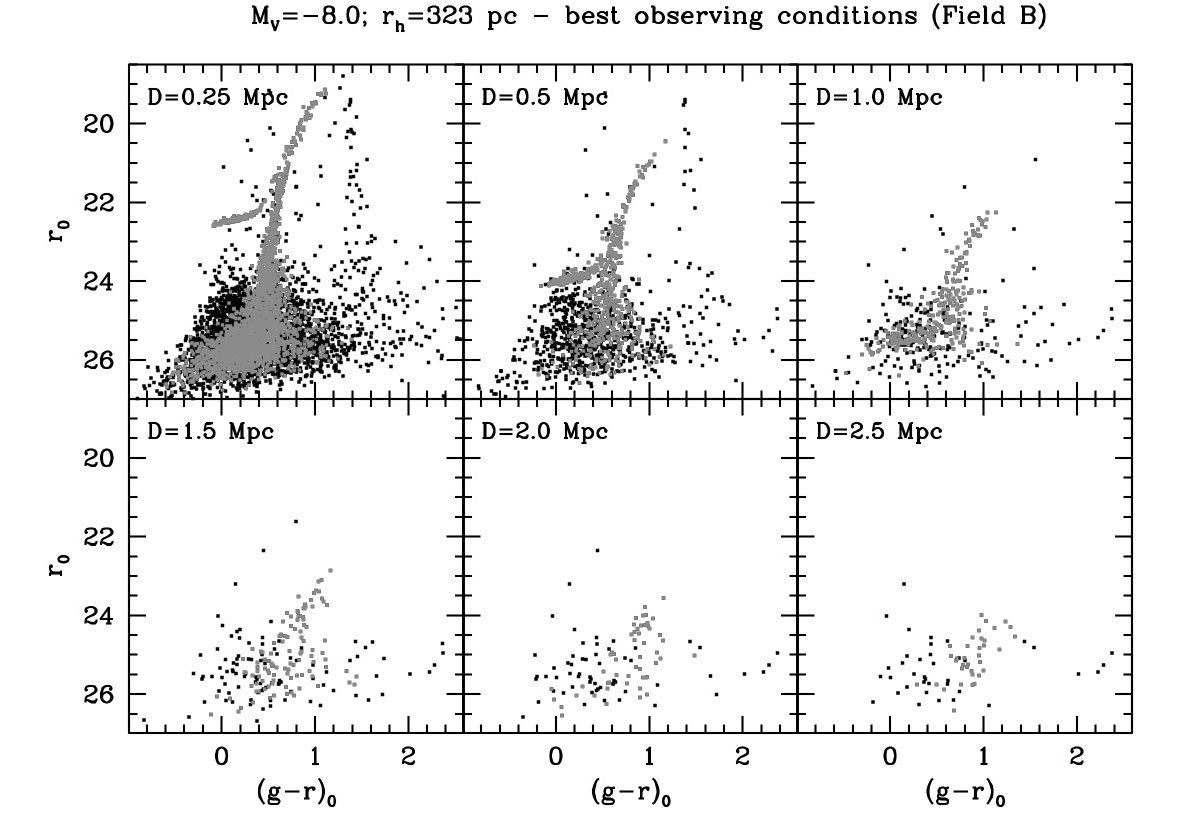}
     \caption{Upper panel: density maps from a $M_V=-8.0$ synthetic galaxy with $R_h=\bar R_h(M_v)$ observed in SECCO (Best case). The galaxy is located at six different distances from D=0.5~Mpc to  D=2.5~Mpc. The density scale is in term of $\sigma$ over the background (see Pap~I). Lower panel: CMD of stars within 1~$R_h$ of the center of the synthetic galaxies shown above. Stars from the synthetic population are plotted in grey while field stars are plotted in black.}
        \label{map8BR0}
    \end{figure}

%%%%%%%%%%%%%%%%%%%%%%%%%%%%%%%%%%%%%%%%%%%%%%%%%%%%%%%%%%%%%%%%%

To illustrate the process of detection of synthetic galaxies with density maps we consider the case of a $M_V=-8.0$ dwarf galaxy. In the upper panel of Fig.~\ref{map8BR0} we show the maps obtained for a $R_h=\bar R_h(M_v)$ model observed in the best conditions, and located at the six distance values adopted for this kind of experiments (see Sect.~\ref{injmap}). It is readily evident that at all the considered distances the synthetic galaxy is detected as an over-density at $\ge 10\sigma$ over the background. While the angular scale varies significantly with the distance, as obvious, there is no doubt that such a stellar system would have been detected in SECCO. 

It may be interesting to note that in the D=0.25~Mpc case the over-density exceeds the limits of the studied field\footnote{That, we recall, is the $17.3\arcmin \times 7.7\arcmin$ field covered by the central LBC chip of each SECCO field.}, still it emerges clearly above the background. The inspection of the corresponding CMD in the lower panel of Fig.~\ref{map8BR0} clearly reveals the presence of an unexpected stellar population, thus confirming the detection beyond any doubt. In the CMDs shown in Fig.~\ref{map8BR0} and in analogous figures below, we have plotted stars of the synthetic galaxies
and of the fore/background population with different colours (grey and black, respectively) for the reader convenience. The distribution of black points provide a view of the typical CMD of the fore/background population in an adjacent ``blank'' field of the same area as those shown in the various panels.

The CMD would allow to unambiguously classify the detected over-densities as local dwarf galaxies (instead of, e.g., clusters/groups of unresolved galaxies, see Pap~I) up to D=1.0~Mpc. In most cases this would also imply that a reliable distance estimate would have been possible with the SECCO observational material, with accuracy depending on the distance, on the total luminosity and on the stellar population of the considered galaxy. In the D=1.5~Mpc case, the CMD does not allow an unambiguous classification and a reliable distance estimate. However it clearly provide support to the detection, strongly suggesting a follow-up of the candidate. On the other hand, the nature of the clean and compact over-densities detected at D=2.0~Mpc and D=2.5~Mpc cannot be established using the CMD, that is too poorly populated and widened by the large photometric errors near the limiting magnitude of the photometry.

%------------------------FIG -----------------------------------
   \begin{figure}
   \centering
   \includegraphics[width=\columnwidth]{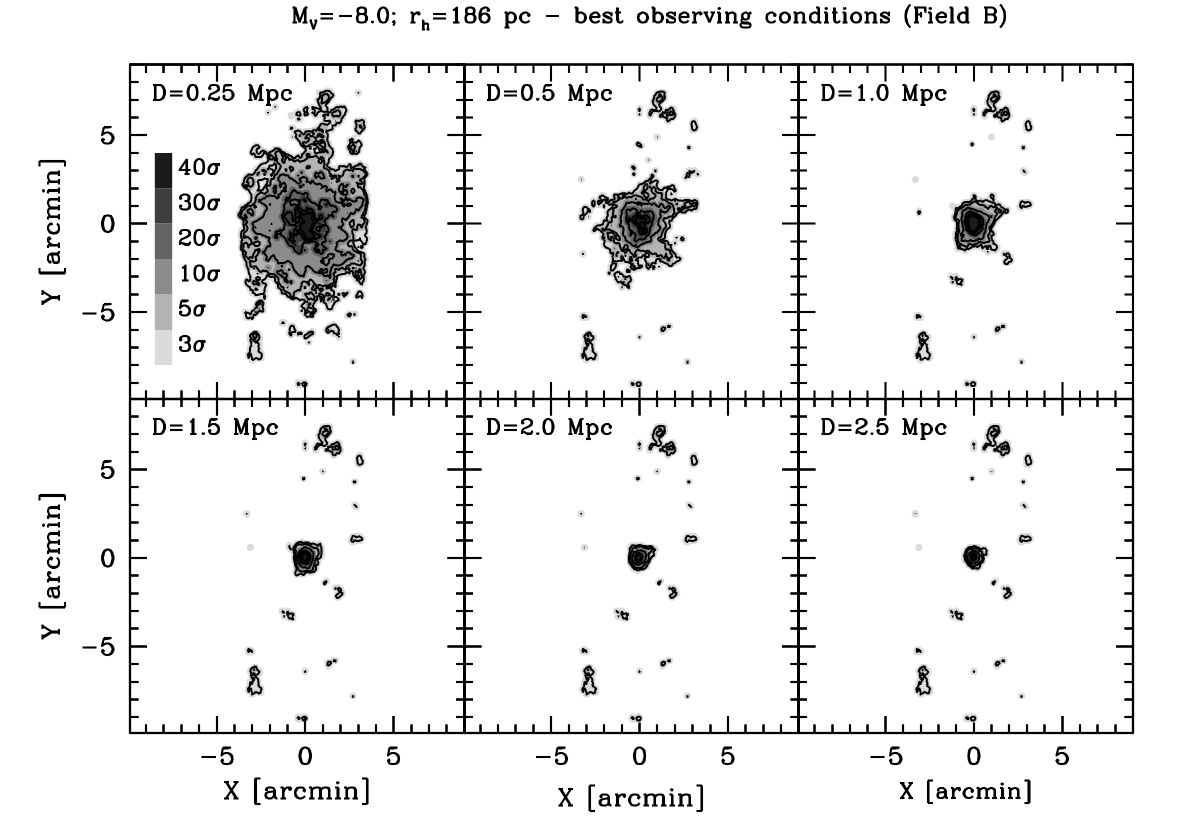}
   \includegraphics[width=\columnwidth]{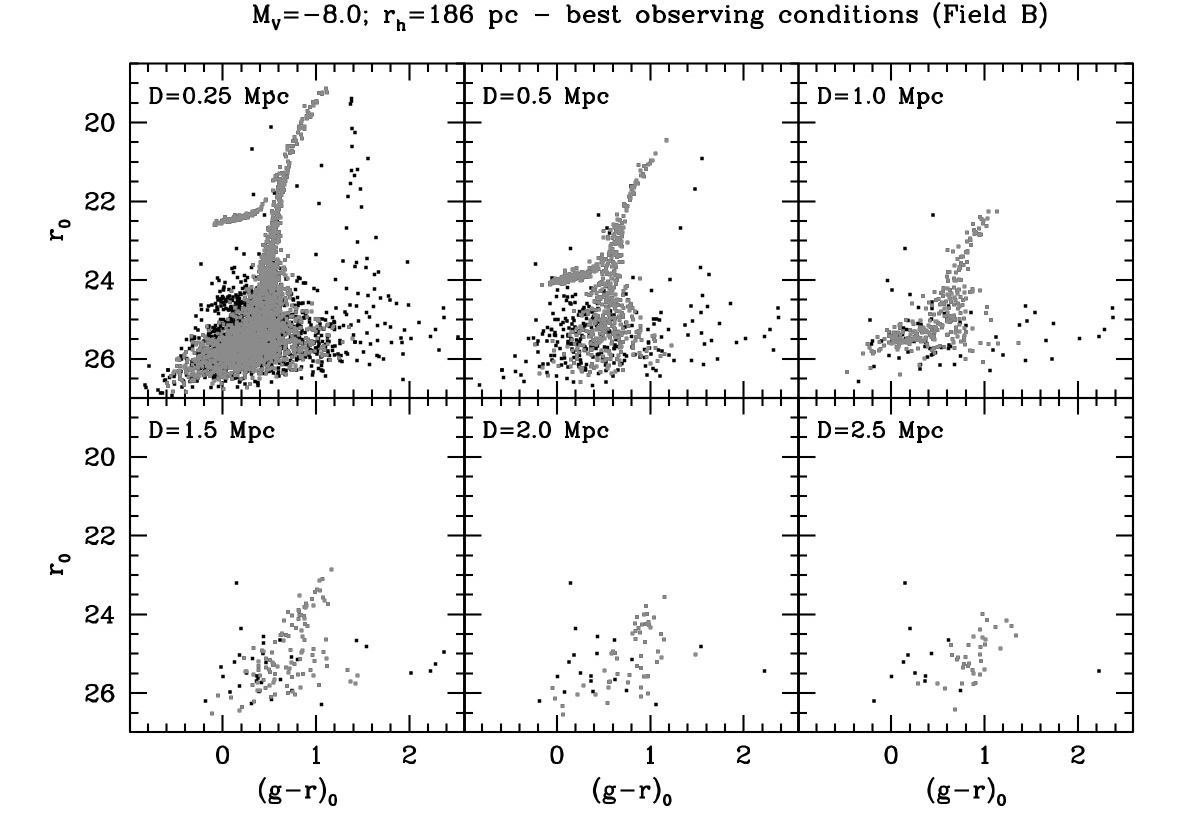}
     \caption{The same as Fig.~\ref{map8BR0} for a $M_V=-8.0$ synthetic galaxy with $R_h=\bar R_h(M_v) -1\sigma_{{\rm log}R_h}$.}
        \label{map8BRm1}
    \end{figure}
%%%%%%%%%%%%%%%%%%%%%%%%%%%%%%%%%%%%%%%%%%%%%%%%%%%%%%%%%%%%%%%%%

%------------------------FIG -----------------------------------
   \begin{figure}
   \centering
   \includegraphics[width=\columnwidth]{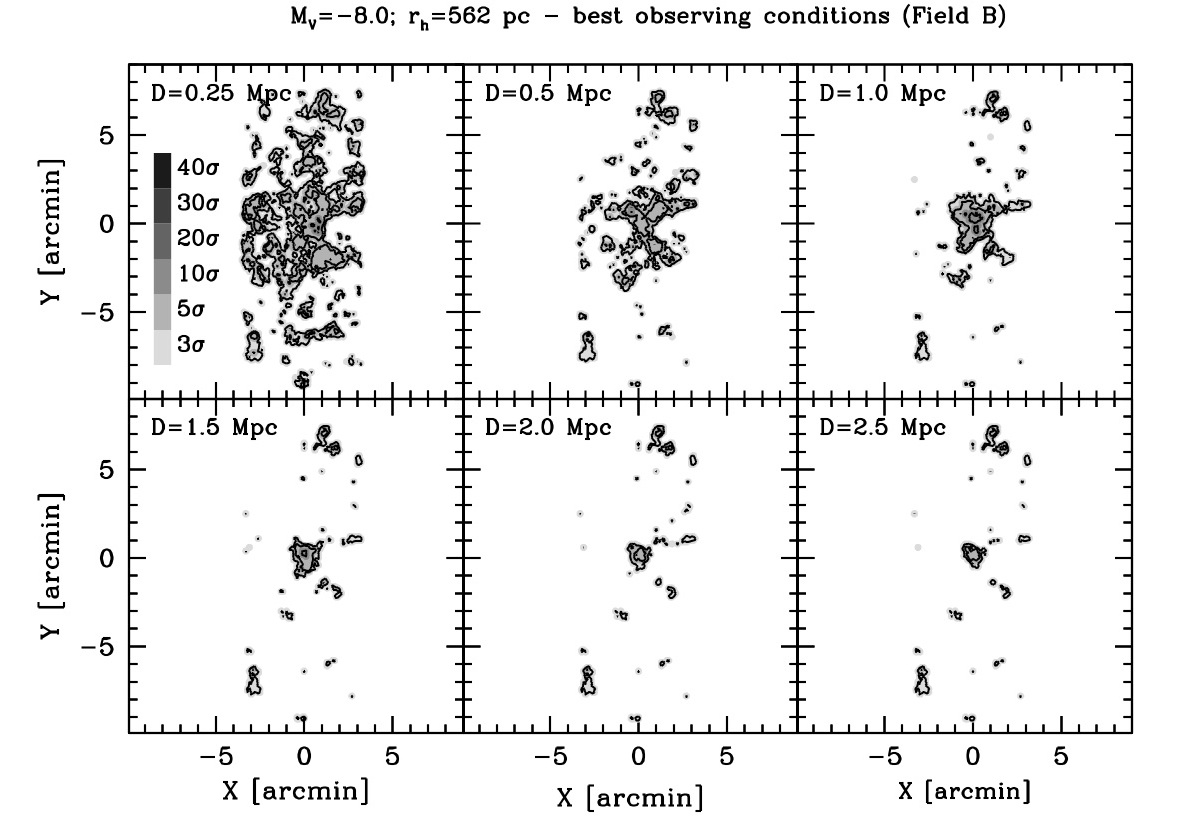}
   \includegraphics[width=\columnwidth]{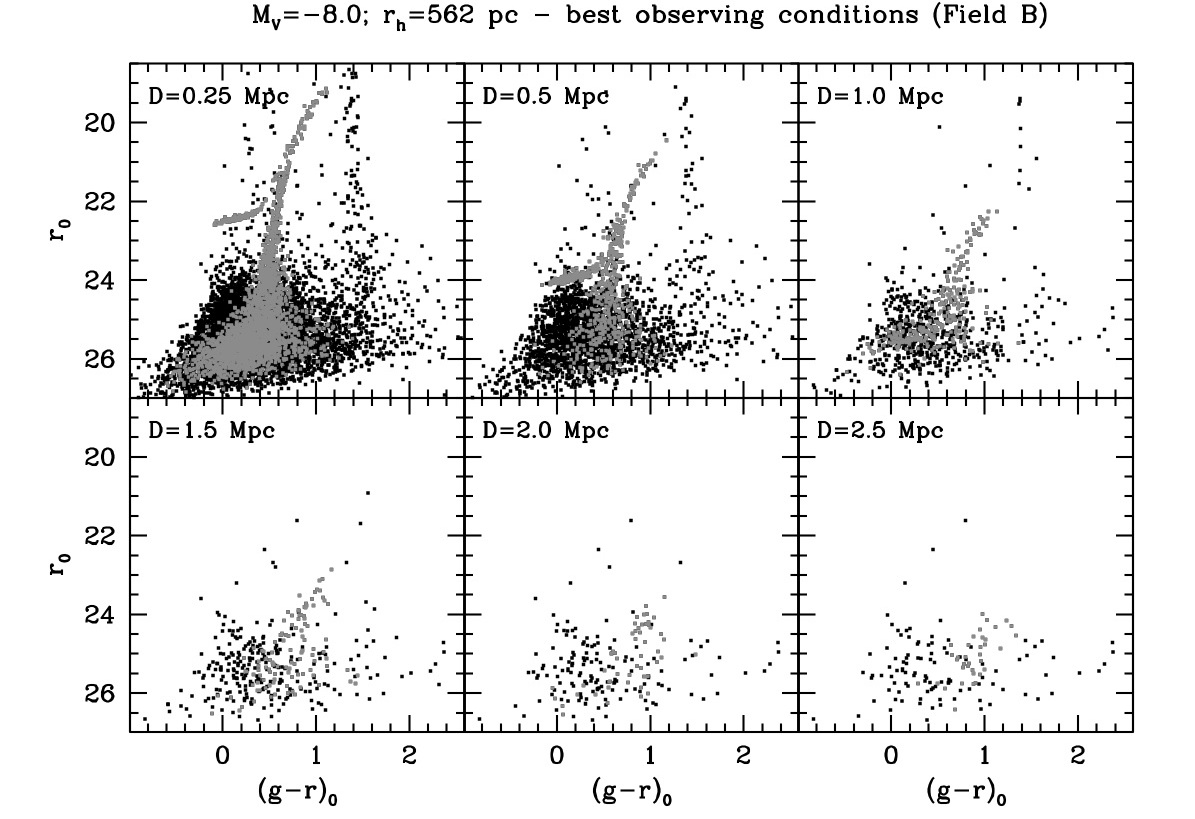}
     \caption{The same as Fig.~\ref{map8BR0} for a $M_V=-8.0$ synthetic galaxy with $R_h=\bar R_h(M_v) +1\sigma_{{\rm log}R_h}$.}
        \label{map8BRp1}
    \end{figure}

%%%%%%%%%%%%%%%%%%%%%%%%%%%%%%%%%%%%%%%%%%%%%%%%%%%%%%%%%%%%%%%%%

Fig.~\ref{map8BRm1} and Fig.~\ref{map8BRp1} shows the same plots as Fig.~\ref{map8BR0} but for models with $R_h=\bar R_h(M_v) -1\sigma_{{\rm log}R_h}$ and with $R_h=\bar R_h(M_v) +1\sigma_{{\rm log}R_h}$, respectively.
The comparison of the maps of the compact, average and diffuse models illustrates the strong impact of the galaxy size on the sensitivity of density maps. As expected, compact galaxies are much more clearly and easily detected.
The $\ge 10\sigma$ over-densities of the $R_h=\bar R_h(M_v)$ case become $\ge 40\sigma$ for the compact model, and in some case reach just $\ge 5\sigma$ for the diffuse model. By definition, the stars from the synthetic galaxies within 1~$R_h$ are the same (for each assumed distance) in all the CMDs for the $M_V=-8.0$ galaxy; what changes in the three cases is the actual extension of $R_h$, leading to the inclusion of more and more field stars in the $R\le R_h$ CMD\footnote{We verified that in our models the angular half-light radius is a good approximation of the radius enclosing the $3\sigma$ iso-density contour, for detected over-densities. Hence, the adoption of $R\le R_h$ for the CMDs shown here is just a convenient, simple and uniform choice that should not seriously bias our ability of confirming/detecting a galaxy from its CMD. Please note that in Pap~I we checked the CMDs over a wide range of radii for {\em all the $\ge 5\sigma$} over-densities detected in our density maps.}. However, {\em in all the considered cases} the detections for $D\le 1.5$~Mpc are clearly confirmed, or strongly supported (for D=1.5~Mpc), by the inspection of the CMD.

%------------------------FIG -----------------------------------
   \begin{figure}
   \centering
   \includegraphics[width=\columnwidth]{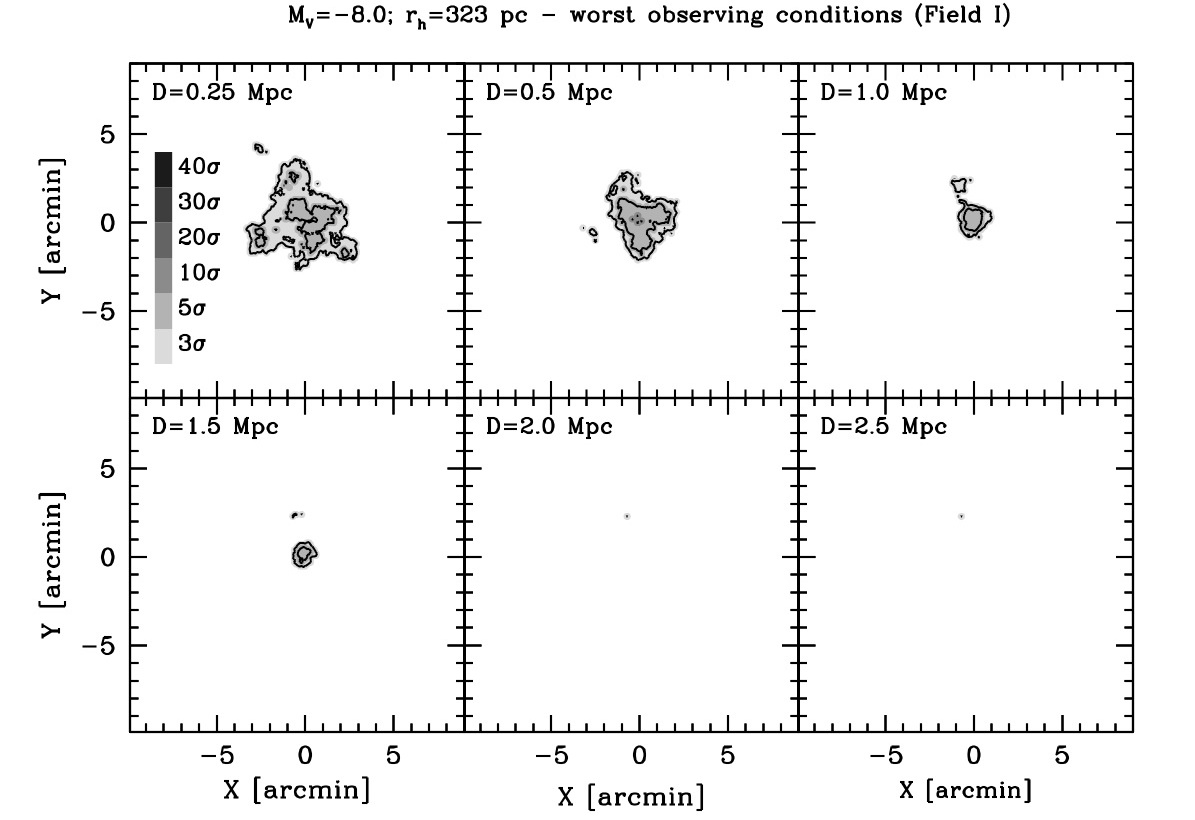}
   \includegraphics[width=\columnwidth]{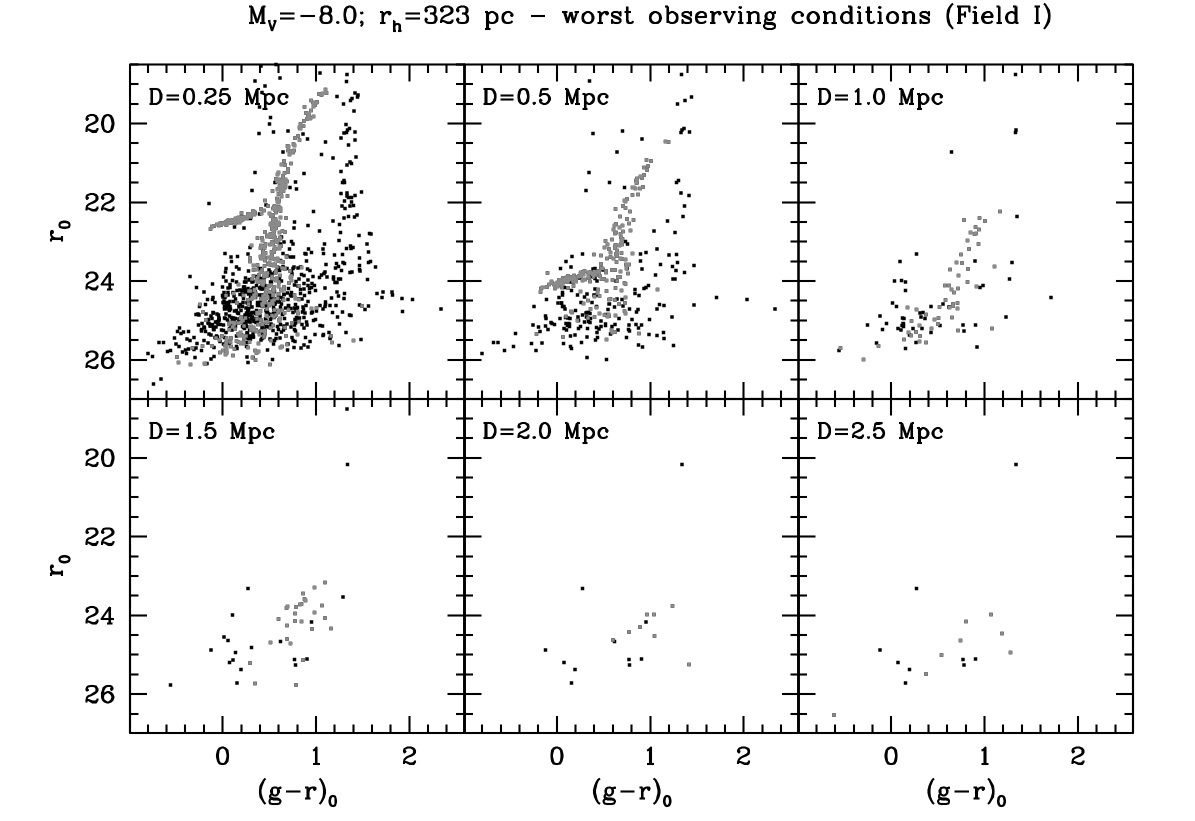}
     \caption{The same as Fig.~\ref{map8BR0} ($R_h=\bar R_h(M_v)$ models) but for observations in SECCO (Worst case).}
        \label{map8IR0}
    \end{figure}

%%%%%%%%%%%%%%%%%%%%%%%%%%%%%%%%%%%%%%%%%%%%%%%%%%%%%%%%%%%%%%%%%

Fig.~\ref{map8IR0} is the strict analogous of Fig.~\ref{map8BR0} but for the case of worst quality observational material in SECCO. It can be readily appreciated that, in this case, a $\ge 5\sigma$ over-density is found only for $D\le 1.5$~Mpc, while more distant galaxies would not be detected. However, in spite of the relatively weak over-density signal, the CMD would provide fairly clear confirmations up to  $D\le 1.0$~Mpc and, possibly, also 
to $D\le 1.5$~Mpc.

\begin{figure}
   \centering
   \includegraphics[width=\columnwidth]{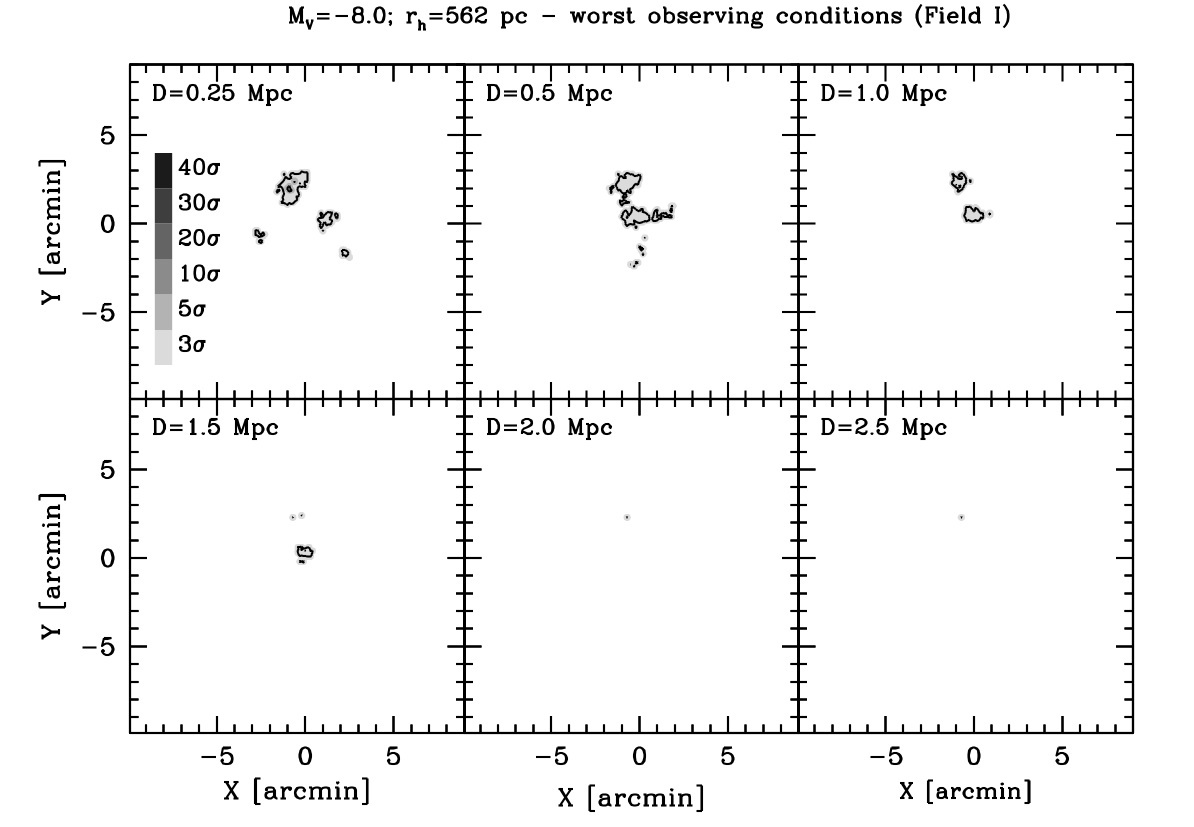}
     \caption{Density maps for the same case as Fig.~\ref{map8IR0} but for a diffuse model.}
        \label{map8IRp1}
    \end{figure}

Finally, Fig.~\ref{map8IRp1} shows that, in the worst observing conditions, a $M_V=-8.0$ dwarf with
$R_h=\bar R_h(M_v) +1\sigma_{{\rm log}R_h}$ would never be detected as an over-density in SECCO. It is worth noting that in some case their presence could be revealed by the inspection of the overall CMD of the associated field (in particular for D$\le 0.5$~Mpc; see, e.g., Fig.~\ref{map8IR0}, lower panel, D=0.25~Mpc case). However, to avoid ambiguity in our definition of sensitivity, in the present study we consider only significant ($\ge 5\sigma$) overdensities as valid detections, and use the CMDs only to support the classification of detected objects as genuine dwarfs.

%------------------------FIG -----------------------------------

%%%%%%%%%%%%%%%%%%%%%%%%%%%%%%%%%%%%%%%%%%%%%%%%%%%%%%%%%%%%%%%%%

The analysis described above for the $M_V=-8.0$ case has been repeated for all the relevant knots of the grid in Table~\ref{lista}. 
%The corresponding density maps and CMDs are collected in Appendix~\ref{app_dens}, below, to preserve the readability of the paper. 

\begin{figure}
   \centering
   \includegraphics[width=\columnwidth]{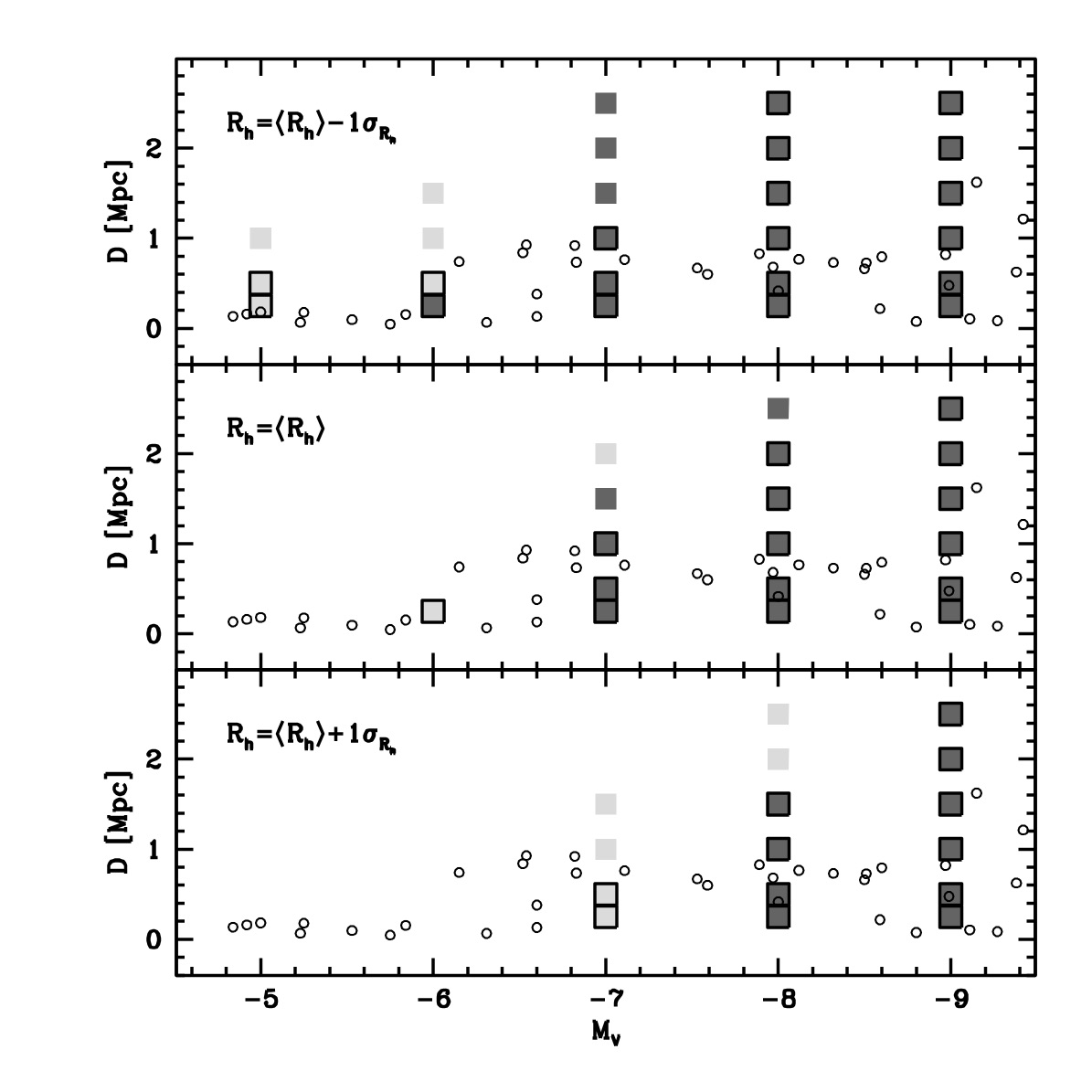}
     \caption{Summary of the density map sensitivity experiments for {\em Best case} SECCO observations. Light-grey squares indicate $\ge 5\sigma$ but $< 10\sigma$ over-density detections, and dark-grey squares correspond to $\ge 10\sigma$ over-density detections; the filled squares that are circled with a dark square correspond to cases where the CMD provides significant support to the detection. Light open circles are dwarf galaxies in the Local Volume in the considered range of $M_V$, from the catalog by \citet{mcc}. The three panels correspond to compact (upper panel), average (middle panel), and diffuse (lower panel) galaxy models.}
        \label{sensB}
    \end{figure}

\begin{figure}
   \centering
   \includegraphics[width=\columnwidth]{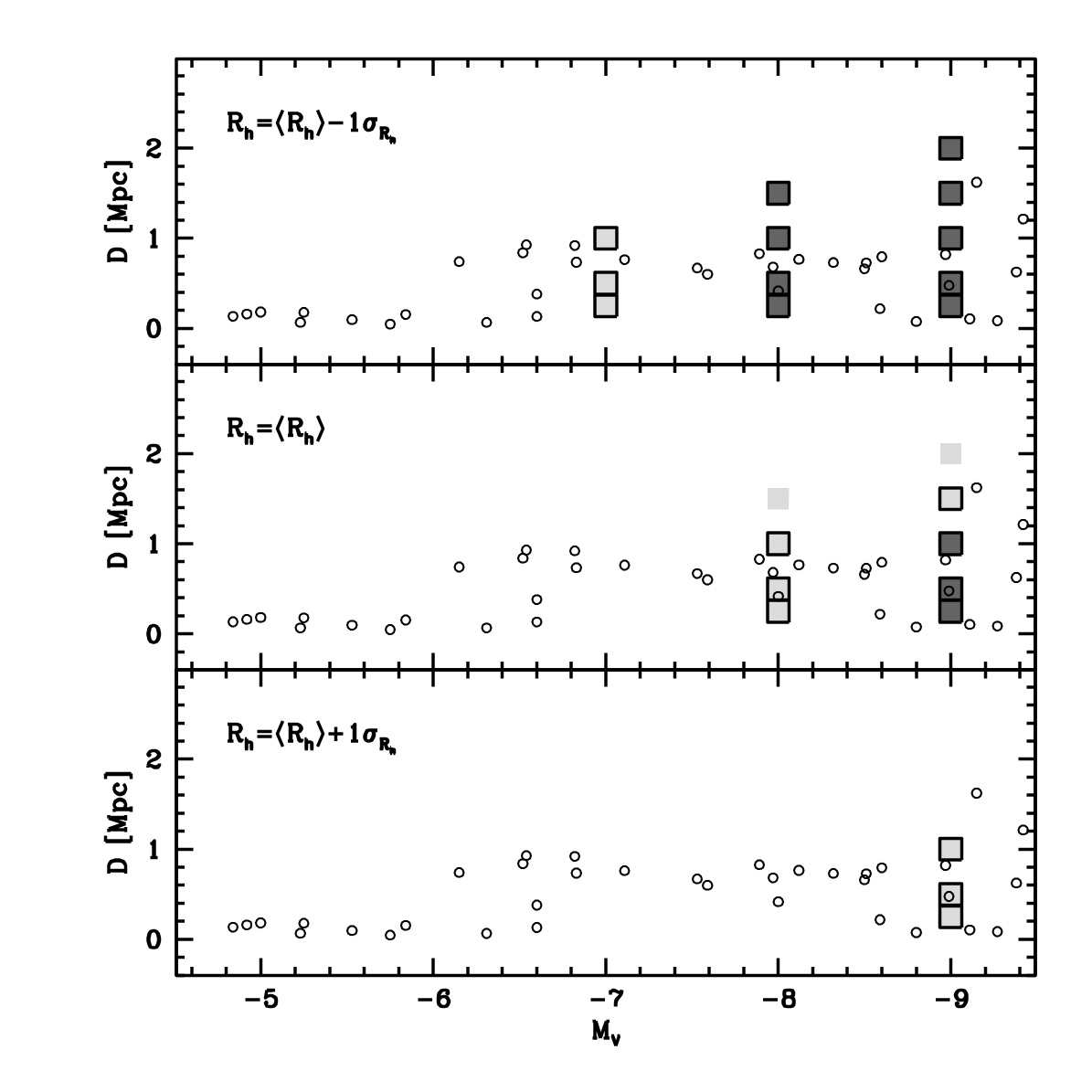}
     \caption{Summary of the density map sensitivity experiments for {\em Worst case} SECCO observations. The meaning of the symbols is the same as in Fig.~\ref{sensB}.}
        \label{sensI}
    \end{figure}

A graphical representation of the main results of this analysis is shown in Fig.~\ref{sensB} and  Fig.~\ref{sensI}.
For any considered case and any adopted distance, we classify the synthetic galaxy and symbolise them in the figures 
according to the following criteria:

\begin{enumerate}

\item it is detected as an over-density at $\ge 10\sigma$ (dark grey solid square)

\item it is detected as an over-density at $\ge 5\sigma$ but $< 10\sigma$ (light grey solid square)

%\item it is not detected as an over-density but it is detected by the inspection of the CMD (open %square); {\bf this occurs only for a handful of the nearest models (D$\le 0.5$~Mpc)}

\item it is not detected at all (not represented)

\end{enumerate}

Moreover we over-plot a black open square on top of the symbols which represent the cases in which
the classification as dwarf galaxy is significantly supported by their CMD.
The results of this analysis are shown in Fig.~\ref{sensB} and  Fig.~\ref{sensI} for Best and Worst case observations, respectively, and are summarised in tabular form in Tab.~\ref{classi}. Fig.~\ref{sensB} shows
that any galaxy with $M_V\le -8.0$ and $D\le 2.5$~Mpc would have been detected by SECCO observations independently of its actual size, in the $\bar R_h(M_v) \pm 1\sigma_{{\rm log}R_h}$ range considered here, if it falls within one of the best-quality SECCO fields, similar to the Best case considered here. In this case, galaxies with $M_V\le -7.0$ are detected out to $D=2.0$~Mpc, if they have $R_h\le \bar R_h(M_v)$, and out to $D=1.5$~Mpc if they are diffuse. In case of compact galaxies, also systems as faint as $M_V=-5.0$ can be detected out to D=1.0~Mpc. In all the panels of this figure we over-plotted the distribution of dwarf galaxies in the Local Volume in the considered range of $M_V$, from the catalog by \citet{mcc}, for reference. We note that, in the considered case, any known dwarf with $M_V\le -6.0$ would have been detected in SECCO. Moreover, the sensitivity of our survey extends to much larger distances than those spanned by known dwarfs in the range $-7.0\le M_V\le -9.0$ (see Sect.~\ref{conc} for further discussion on this point). 

%------------------------FIG -----------------------------------

%%%%%%%%%%%%%%%%%%%%%%%%%%%%%%%%%%%%%%%%%%%%%%%%%%%%%%%%%%%%%%%%%

%------------------------FIG -----------------------------------

%%%%%%%%%%%%%%%%%%%%%%%%%%%%%%%%%%%%%%%%%%%%%%%%%%%%%%%%%%%%%%%%%

In worst-quality data (Fig.~\ref{sensI}) we will still be able to detect any dwarf with $R_h\le \bar R_h(M_v)$ out to $D=2.0$~Mpc for $M_V=-9.0$ and out to $D=1.5$~Mpc for $M_V=-8.0$. Compact dwarfs are seen out to $D=1.0$~Mpc also for $M_V=-7$, while only the brightest and most nearby {\em diffuse} galaxies can be detected.
We will show below that for $M_V\le -10.0$ even diffuse galaxies would have been detected by visual inspection in SECCO images, independently of the considered dataset. 

It is worth noting that in the majority of the considered cases the nature of the detected (or undetected) over-density can be ascertained from the inspection of its CMD. Finally we note that, within the considered grid of models, the lowest central surface brightness corresponding to a $\ge 5\sigma$ over-density is $\mu_{V,h}= 30.0$~mag/arcsec$^2$.

\section{Sensitivity of the visual inspections}
\label{sensima}

In this section we provide a basic exploration of the process of search of stellar counterparts by visual inspection of the images. This is mainly intended (a) to assess the luminosity limit above which a typical dwarf cannot be missed on our image, (b) to provide general guidelines on what we can expect to detect in our images, and (c) to provide a framework to interpret actual detections, obtaining rough constrains on the characteristic of a detected system by comparison with images of synthetic galaxies.

\begin{figure}
   \centering
   \includegraphics[width=\columnwidth]{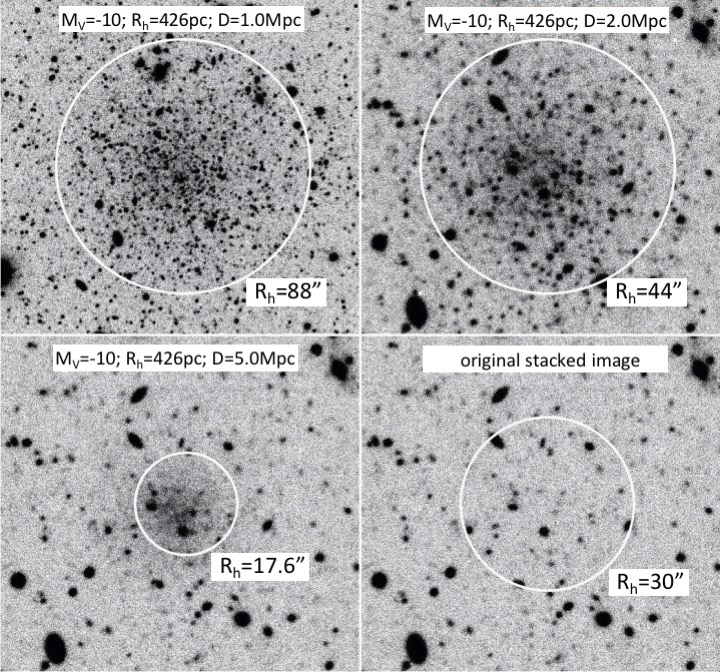}
     \caption{Synthetic dwarf galaxies of $M_V=-10.0$ added to SECCO {\em Worst case} images, for different assumptions of the distance. 
     All the stamp images are centred on the center of the synthetic galaxy; a circle with radius =$R_h$ is over-plotted.
     The stamp image in the lower right corner is the original image, without synthetic galaxy, that is shown for reference with a conventional circle of radius $=30\arcsec$.}
        \label{ima10}
    \end{figure}

In Fig.~\ref{ima10} we show few examples of portions of {\em Worst case} images centred on a synthetic dwarf galaxy that was added as described in Sect.~\ref{injima}. We show these images here as reference to help the reader to have a feeling of the quality to the adopted datasets and of the realism of the simulations.  
A set of figures analogous to Fig.~\ref{ima10} are presented in Appendix~\ref{app_ima} for the most relevant knots of the grid of considered models (Table~\ref{lista}). 
%While the presented set is not very extended we feel that can provide a useful guideline to evaluate the appearance of faint and low surface brightness local dwarfs in images from 8m class telescopes.  

The first three panels of Fig.~\ref{ima10} (from left to right and from top to bottom), show images of a synthetic dwarf galaxy with $M_V=-10.0$ and $R_h=\bar R_h(M_v)$ located at the distances 
of 1, 2 and 5 Mpc, respectively. The fourth (lower right corner) shows the original ``empty'' image, for reference. A circle of radius equal to the half-light radius is over-plotted, for reference, and its value in arcsec is also reported.

%%%%%%%%%%%%%%%%%%%%%%%%%%%%%%%%%%%%%%%%%%%%%%%%%%%%%%%%%%%%%%%%%

%%------------------------FIG -----------------------------------
%   \begin{figure}
%   \centering
%   \includegraphics[width=\columnwidth]{Mv10R0imaI.jpg}
%   \includegraphics[width=\columnwidth]{Mv10Rp1imaI.jpg}
%     \caption{Synthetic dwarf galaxies of $M_V=-10.0$ added to SECCO Field~I images, for different assumptions of the distance. The upper set of stamp images refers to average models, the lower set to diffuse models. All the stamp images are centred on the center of the synthetic galaxy; a circle with radius =$R_h$ is overplotted.
%     The scale of the last four images of each set have the same scale, while the first two have been zoomed out to include the apparent half-light radius of the considered model. The stamp image in the lower right corner of each set is the original image, without synthetic galaxy, that is shown for reference with a conventional circle of radius $=30\arcsec$.}
%        \label{ima10}
%    \end{figure}
%

%
%%%%%%%%%%%%%%%%%%%%%%%%%%%%%%%%%%%%%%%%%%%%%%%%%%%%%%%%%%%%%%%%%

%As mentioned in Sect.~\ref{injima}, we produced stacked images for each of the simulated dwarf %galaxies adopted to assess the sensitivity of our
%study using the density maps. 
In the case of the visual inspection, the examination of the stacked images including synthetic dwarfs allowed us to draw the following conclusions: 

\begin{itemize}

\item $M_V\le -10.0$ dwarfs out to D$=5$~Mpc would have been detected in SECCO by visual inspection, independently of the quality of the available observational material and of the galaxy size, within $R_h=\bar R_h(M_v) \pm 1\sigma_{{\rm log}R_h}$. As unassuming as it may appear (Fig.~\ref{imaI10}), even the diffuse models are clearly noticed when the images are carefully inspected as we did for real SECCO images (Pap~I; see Sect.~\ref{lsb}, for real examples). These very low surface brightness galaxies are pretty evident when a simple gaussian smoothing of a few pixels is applied to the images. 

\item the effect of galaxy size is even stronger in images than in density maps. In fact, it turns out that no diffuse model is detected by visual inspection for $M_V\ge -9.0$; in the following of this section we will not deal anymore with diffuse models.

\item in our survey, galaxies can be partially resolved up to $D\simeq 3.0$~Mpc. As unresolved systems they can be detected out to relatively large distances (see, again, Sect.~\ref{lsb}) . 

\end{itemize}

%------------------------FIG -----------------------------------
   \begin{figure}
   \centering
   \includegraphics[width=\columnwidth]{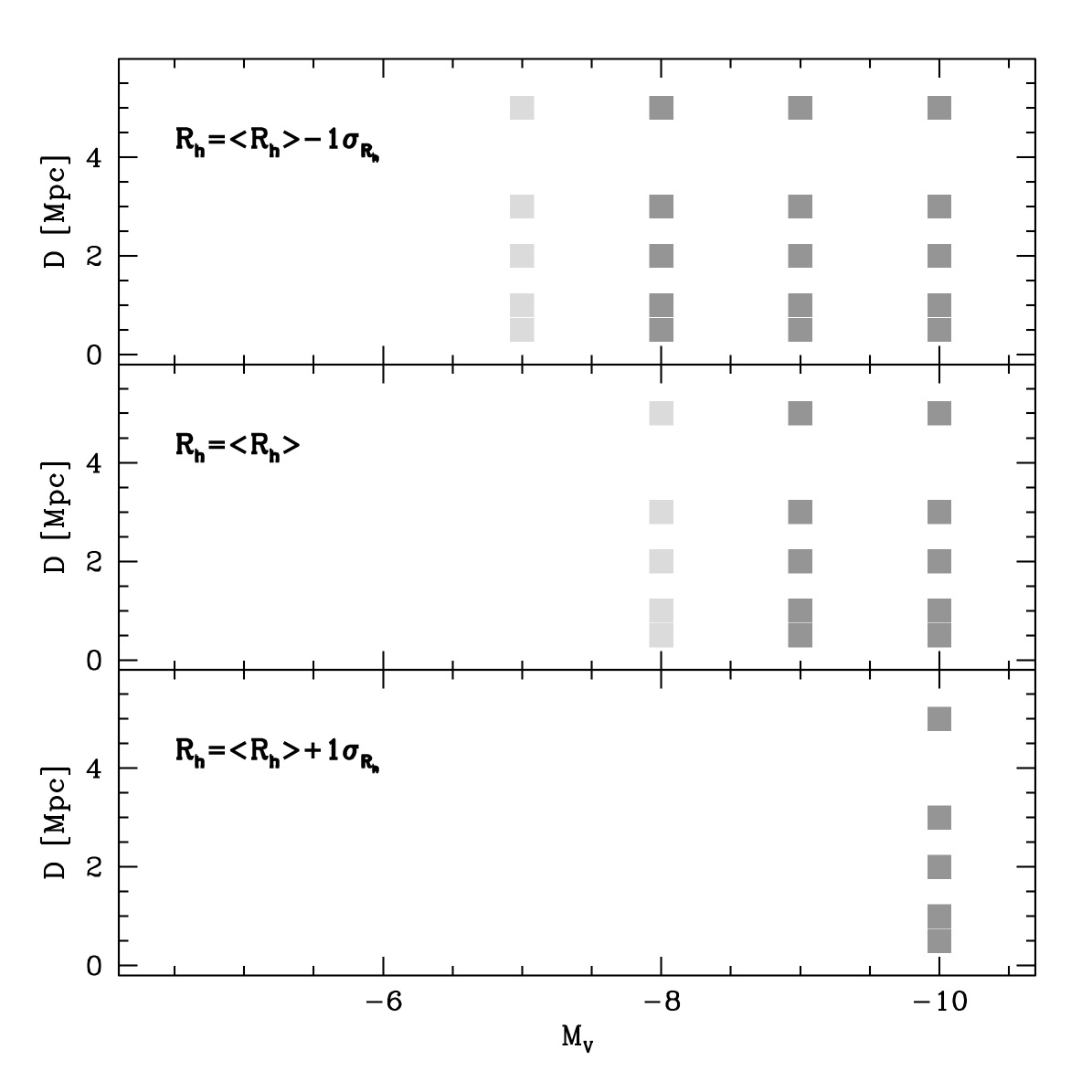}
   \includegraphics[width=\columnwidth]{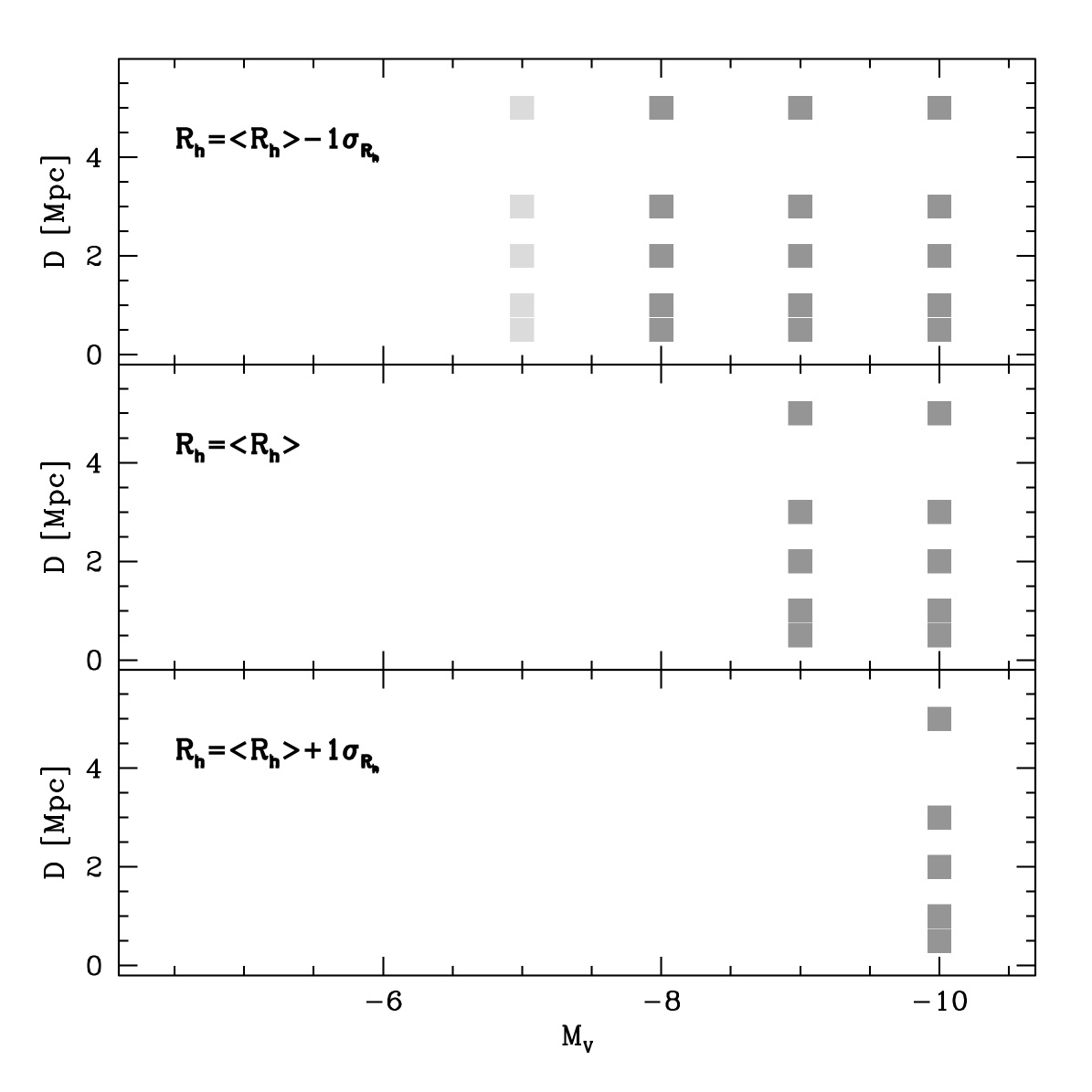}
     \caption{Summary of the visual inspection sensitivity experiments for {\em Best case} (upper panels) and {\em Worst case} images. Dark and light grey squares indicate a clear and doubtful detection, respectively.}
        \label{imasens}
    \end{figure}

%%%%%%%%%%%%%%%%%%%%%%%%%%%%%%%%%%%%%%%%%%%%%%%%%%%%%%%%%%%%%%%%%

The results of the critical analysis of the entire set of images with synthetic galaxies is summarised in Fig.~\ref{imasens}. It is clear that visual inspection is a useful complement to the density map analysis but does not significantly enhance the sensitivity of the survey in the range of distances that is of primary interest for SECCO, i.e  0.25~Mpc$\le D\le 2.0$~Mpc (see Pap~I). 

On the other hand it extends the sensitivity of the survey to larger distances, especially for compact models. 
Once again, the example of SECCO~1 (at $D\simeq 16.5$~Mpc) is fairly illustrative: the curious configuration of a few blue stars triggered the attention during the visual inspection of the images but was largely below the detection threshold in density maps (Pap~I).  In Sect.~\ref{lsb}, below, we empirically demonstrate our ability to spot unassuming spheroidal dwarfs as faint as $M_V\sim -8.0$ at the distance of the Virgo cluster, independently of their position in the field.
 
The synthetic galaxy with the lowest surface brightness that we detect by visual inspection has $\mu_{V,h}= 28.1$~mag/arcsec$^2$, almost two magnitude brighter than the limit achieved with density maps. This can be taken as the fundamental sensitivity limit for visual inspection of SECCO 
images\footnote{With a simple rescaling on aperture photometry we obtained rough estimates of the magnitude and surface brightness of the faintest among the Virgo dwarf spheroidal presented in Sect.~\ref{lsb}, VC1 and VC2. We find $M_V\sim -7.0$ and $\mu_{V,H}\sim 27.8$~mag/arcsec$^2$ (VC1), fully consistent with the limits from synthetic dwarfs.}. 

We are aware that the simple approach adopted for this part of the analysis may suffer from {significant bias in favour of detection, especially for faint dwarfs ($M_V< -9.0$). A bias may be caused, for example, by placing the synthetic galaxy near the center of the field and by avoiding the superposition with heavily saturated stars (see Sect.~\ref{injima}). To fully reproduce the actual process of visual inspection performed on real data we should have produced a large number of images in which synthetic galaxies are added at random position within the field and/or not added at all; then we should have these images independently inspected by two of us, as done in Pap~I. However, such an expensive approach would not add any significant piece of information with respect to the main goal of the overall analysis, since, as said, density maps have a much higher sensitivity than visual inspection in the crucial region of the parameter space. On the other hand the basic procedure described in this section is adequate to get a quantitative idea of the visibility of dwarfs at larger distances than those sampled by the density maps. Finally, in Fig.~\ref{imasens} we distinguish between cases in which the galaxy cannot be missed and less obvious detections, thus providing additional guidance on this aspect.

\section{An empirical test: Low Surface Brightness dwarfs in Virgo}
\label{lsb}

During the visual inspection of SECCO images (Pap~I) we identified seven fluffy and roundish unresolved Low Surface Brightness (LSB) galaxies with typical diameters of $20\arcsec - 40\arcsec$. 
Six of  them lie in three of the four (over 25) SECCO fields that are projected within the wide boundaries of the Virgo cluster of galaxies \citep{extvirgo}. The Virgo cluster is known to host a large population of faint LSB galaxies that can be classified as dwarf spheroidals \citep[dSphs, see, e.g.,][and references therein]{phill,sabatini,cald_virgo}. It is likely that a large fraction of this population is still to be discovered; the ongoing Next Generation Virgo Survey \citep[NGVS,][]{lauraf} is expected to provide a fundamental contribution in this respect \citep[see, e.g.][]{davlsb_virgo,mihos}.
Hence it is reasonable to assume that the LSB galaxies we discovered in SECCO are physically associated to the Virgo cluster. 

We decide to briefly report on them here for two main reasons. First they are interesting on their own, since they appear among the faintest and lowest surface brightness dwarfs ever discovered in Virgo, also showing intriguing clustering properties and, second and most relevant for the present paper, they provide real examples of how faint  LSB dwarfs have been {\em actually found} by visual inspection in SECCO, thus giving independent support to our sensitivity analysis based on synthetic galaxies. Finally, three of them have been found in the same images where SECCO~1 (also likely lying in Virgo) has been found \citep[Field~D, see Pap~I and][]{secco_l1} thus providing the basis for a fruitful comparison with that stellar system, that appears peculiar under various aspects. The seven LSB systems are listed in Table~\ref{lsb_tab}, where the adopted naming convention is also defined. In the following we will drop the ``SECCO-LSB'' suffix for brevity. None of the listed systems is found in the Simbad\footnote{\tt http://simbad.u-strasbg.fr/simbad/} or NED\footnote{\tt http://ned.ipac.caltech.edu} astronomical databases, nor in the recent catalogs by \citet{extvirgo} and \citet[][with one exception, see Tab.~\ref{lsb_tab}]{davlsb_virgo}\footnote{Some of them are clearly seen also in NGVS images when inspected through the web interface {\tt \tiny www4.cadc-ccda.hia-iha.nrc-cnrc.gc.ca/en/megapipe/access/graph.html}}.

%%%%%%%%%%%%%%%%%%%%%%%%%%
%
\begin{table*}
  \begin{center}
  \caption{Candidate Low Surface Brightness dwarfs in the Virgo cluster.}
  \label{lsb_tab}
  \begin{tabular}{lccccccc}
ID    &  RA$_{J2000}$    &  Dec$_{J2000}$   &  g     &  r   & $R_h$   & $\mu_{V,h}$ & D$_{M87}$\\
      & [deg]  & [deg]  & [mag]  & [mag] &[arcsec] & [mag/arcsec$^2$] & [deg] \\
\hline
SECCO-LSB-VC1 & 182.26566 &  +4.62742 &  &   &  &  & 9.4 \\
SECCO-LSB-VC2 & 182.26848 &  +4.61088 &  &   &  &  & 9.4 \\
SECCO-LSB-VD1 & 185.28652 & +13.59057 & $\sim19.5$ & $\sim18.1$  & $\sim13.2$ & $\sim26.3$ & 2.6 \\
SECCO-LSB-VD2 & 185.28589 & +13.57372 & $\sim23.2^a$ & $\sim22.6^a$  &  $\sim2.5^a$ & $\sim26.8^a$ & 2.6 \\
SECCO-LSB-VD3 & 185.34763 & +13.58349 & $\sim20.6$ & $\sim19.8$  &  $\sim5.8$ & $\sim25.9$ & 2.6 \\
SECCO-LSB-U1$^c$  & 167.39125 &  +5.31381 &  &   &  & & 21.3 \\
SECCO-LSB-VX1 & 188.48816 &  +8.40119 &  &   &  &  & 4.1 \\
\hline
\end{tabular} 
\tablefoot{The last group of letters in the ID (after SECCO-LSB-, whose meaning is obvious) may contain a V, indicating the likely association with Virgo, and then the letter of the SECCO field where the galaxy has been found (in this case Field~C, Field~D and Field~X;see Pap~I), and finally a number to distinguish between dwarfs lying in the same SECCO Field.
D$_{M87}$ is the angular distance from the M~87 galaxy, here taken as a proxy for the center of the Virgo cluster.\\ 
Integrated magnitudes and surface brightness values have been corrected for reddening assuming E(B-V)=0.048 (Pap~I).
\\
$^a$ Particularly uncertain because of the faintness of the object.\\
$^b$ Same coordinates, within $1.0\arcsec$, of LSBVCC357 listed by \citet{davlsb_virgo}.
$^c$ This galaxy lies $\sim 4\degr$ beyond the zero velocity boundary of Virgo as estimated by \citet{naso}, between 
the Leo~I and Leo~II galaxy groups, both lying at a distance of about 10-12~Mpc from us, according to the NED database.}
\end{center}
\end{table*}
%%%%%%%%%%%%%%%%%%%%%%%%%%%%%%%%%%%%%%%%%%%%%%%%%%%%%%%%%%%%%%%%%%%%%%%%%%%%%%%%%%%%%%%%%%%%%%%%%%%%%%%%%%%%

%%%%%%%%%%%%%%%%%%%%%%%%%%%%%%%%%%%%%%%%%%%%%%%%%%%%%%%%%%%%%%%%%

%------------------------FIG -----------------------------------
   \begin{figure}
   \centering
   \includegraphics[width=\columnwidth]{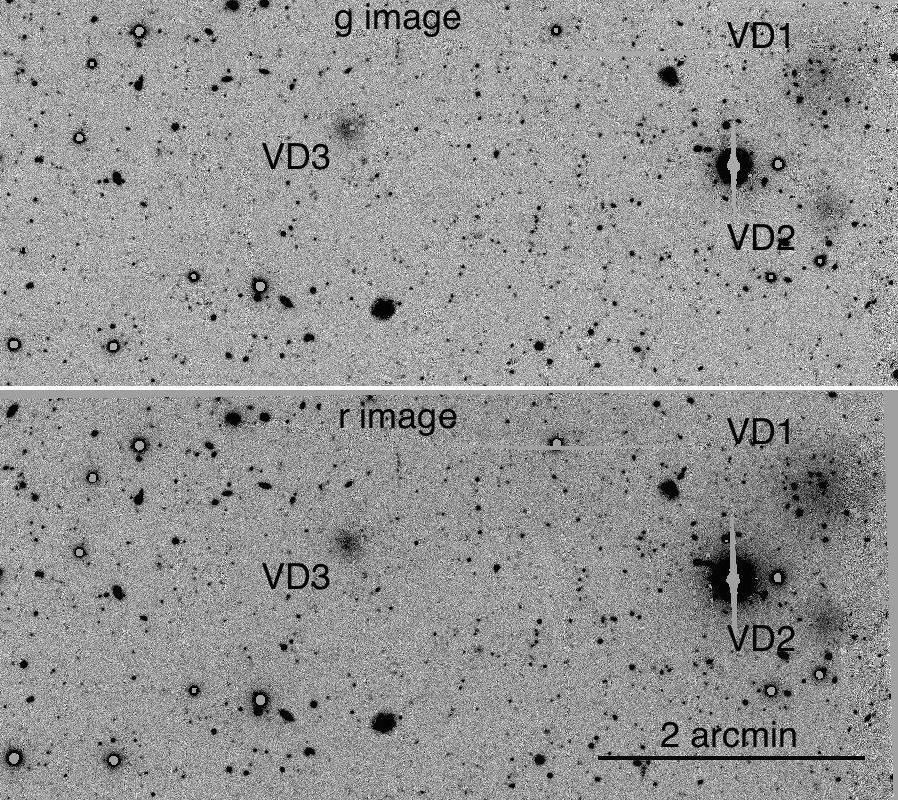}
     \caption{A portion of Field~D g-band (upper panel) and r-band (lower panel) images enclosing
     the three LSB galaxies identified in this field. North is up, East to the left.}
        \label{lsb1}
    \end{figure}

%%%%%%%%%%%%%%%%%%%%%%%%%%%%%%%%%%%%%%%%%%%%%%%%%%%%%%%%%%%%%%%%%

Fig.~\ref{lsb1} displays the three LSB dwarfs found in Field~D in a single zoomed image. This is possible because they are remarkably close together. Of the six candidate dSphs that we find in the 2131 arcmin$^2$ covered by the four SECCO fields sampling Virgo, (these) three lie within a circle of area $\simeq 11$~arcmin$^2$, hardly consistent with a random distribution, thus suggesting a physical association (see below for the case of VC1 and VC2). At the distance of 16.5~Mpc 
\citep[adopted as the mean distance to Virgo,][]{mei}, the projected separation between the centres of VD1 and VD2 is 4.8~kpc, a distance comparable with the sum of their apparent diameters; VD3 is just 17.5~kpc from VD1. While inspecting NGVS images we noted another similar LSB dwarf lying just beyond the northern limit of the SECCO image of Field~D, at $\simeq 5.2\arcmin\simeq 25.0$~kpc from VD1, in the approximate position (RA, Dec)=($185.310\deg$, $13.675\deg$)).

We used GALFIT \citep{galfit} to estimate magnitudes, radii and central surface brightness of the LSB dwarfs lying in Field~D. We limit this analysis to VD1, VD2 and VD3 because the are the most prominent dwarfs of the sample and to have quantitative benchmarks to compare with SECCO~1, from exactly the same images. We fitted Sersic models to the g and r images. At a first pass we leave also the Sersic index $n_s$ as a free parameter, then we repeated the fit keeping $n_s$ fixed at the best-fit value of the first pass. All the three galaxies have $n_s\le 1.0$, as typical of faint dwarfs in Virgo \citep{cote}. Their integrated magnitudes, half-light radii and central surface brightness, as derived with GALFIT, are listed in Table~\ref{lsb_tab}. We remark here that we made no attempt to mask background sources, hence the reported photometry may suffer from some contamination \citep[in particular VD1 that overlaps with a compact group of galaxies in the background, SDSSCGB22677,][]{mc09}. Moreover VD1 and VD2 lie near a corner of Field~D and are partially out of one of the two g+r pairs of SECCO images of Field~D, and the adjacent bright star provide a higher than average and spatially variable background. Hence the obtained surface
photometry is not optimal. From the comparison of the independent measures on two images per passband for VD3, keeping a conservative attitude, we conclude that the typical uncertainty on integrated magnitudes and central SB are $\la \pm 0.3$~mag and $\la$ 20\% on $R_h$. When D=16.5~Mpc is assumed, VD1, VD2 and VD3 have $M_V=$~-12.4, -10.9, and -8.2, respectively, and they fit nicely into the $M_V$~vs.~$R_h$ and $M_V$~vs.~$\mu_{V,0}$ relations defined by local dwarfs \citep{mcc}.

%------------------------FIG -----------------------------------
   \begin{figure}
   \centering
   \includegraphics[width=\columnwidth]{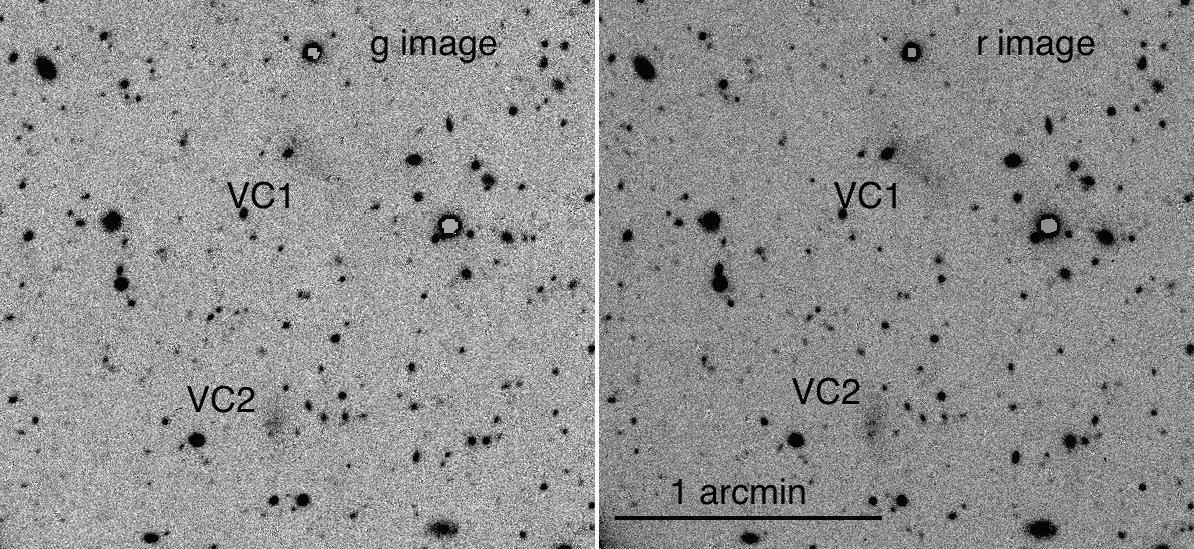}
   \includegraphics[width=\columnwidth]{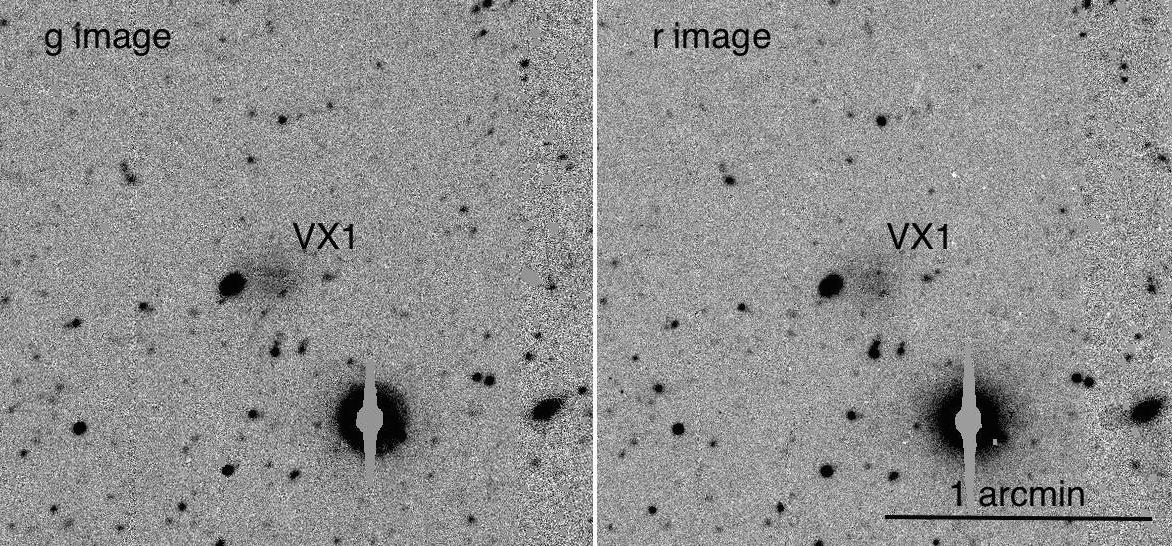}
   \includegraphics[width=\columnwidth]{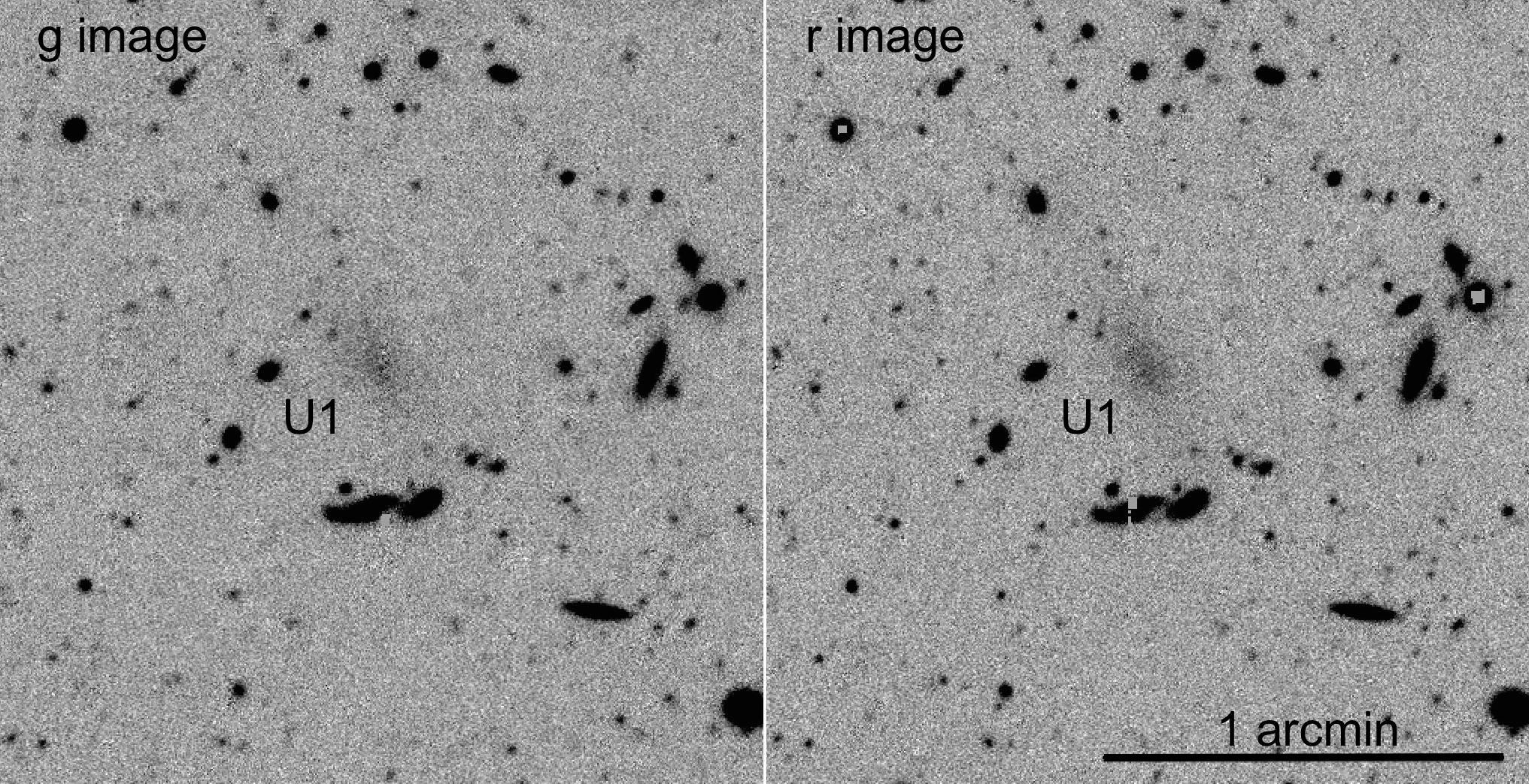}
     \caption{g-band (left panels) and r-band (right panels) stamp images of the LSB galaxies found in Field~C
     (upper pair of images), Field~X (middle pair of panels), and Field~U (lower pair of panels). North is up, East to the left.}
        \label{lsb2}
    \end{figure}

%%%%%%%%%%%%%%%%%%%%%%%%%%%%%%%%%%%%%%%%%%%%%%%%%%%%%%%%%%%%%%%%%

The g and r band images of the remaining LSB dwarfs listed in Tab.~\ref{lsb_tab} are shown in Fig.~\ref{lsb2}.
It is worth noting that, similarly to the case of Field~D,  also the two objects identified in Field~C are very close one another. VC1 and VC2 are separated by just $1.0\arcmin$ in the plane of the sky, corresponding to $\simeq 4.8$~kpc at the distance of Virgo. Our sample of Virgo LSB is indeed too scanty to draw any general conclusion, but it is quite intriguing to note that five of the six galaxies more certainly attributable to Virgo are seen in very tight groups of two and three dwarfs. 

The LSB dwarf found in Field~U is $\sim 4\degr$ beyond the zero-velocity contour of the Virgo cluster derived by \citet[][$\sim 17\degr$]{naso}, hence it cannot be considered as a member of Virgo. However it is located, in the plane of the sky, in the middle between two galaxy groups (Leo~I and Leo~II/NGC~3607) that lie at similar distance; if it is associated with such groups, as it seems likely, also U1 is a dwarf galaxy similar to the other ones listed in Table~\ref{lsb_tab}.

The detection of the faint LSB dwarfs discussed in this section fully support the results presented above, providing an independent empirical validation of the search by visual inspection we performed in Pap~I and of the quantitative sensitivity analysis presented here.

\subsection{Comparison with SECCO~1}

Figure~\ref{secco1_vd} allows a direct comparison of the size, brightness, morphology and colours of SECCO~1 and of the LSB dwarfs identified in the same SECCO images (Field~D). The (reasonable) assumption that all the considered stellar systems lie within the Virgo cluster makes the comparison especially insightful.
On the left panel of the image we have highlighted the main body of SECCO~1 \citep{secco_l1} as well as the possible secondary body identified by \citet{sand15}. Here we have highlighted with an ellipse a possible additional grouping of blue stars that may be associated with the system; a similar swarm of faint blue stars is also seen just to the east of the secondary body. If these structures are indeed physically associated, then SECCO~1 is a system even more complex and anomalous than already believed \citep{adams15}, and a tidal \citep{duc} or ram-pressure \citep{fuma,yoshi} stripping phenomenon would gain support as an hypothesis for its origin \citep[see][]{secco_l1}.

The comparison with the Virgo dSphs lying in the same image, in particular with the faintest  and most diffuse one, VD2, lead to some interesting conclusions:

\begin{itemize}

\item any diffuse component associated to SECCO~1 should be {\em significantly fainter} than the central SB of VD2, i.e. $\mu_{V}<< 26.5$~mag~arcsec$^2$. The comparison with our simulated images consistently indicates that it should be in fact $\mu_{V}<< 27.0$~mag~arcsec$^2$

\item any diffuse component associated to SECCO~1 is {\em significantly bluer} than the dSphs. Indeed some very feeble
fuzzy blue light is seen within the circle enclosing the main body, especially to the east of the brightest sources, and in the middle of the secondary body.

\end{itemize} 

%The above comparison suggests that the old stellar halo that is ubiquitously found even in the most %actively star-forming galaxies \citep{tht} should be a very minor component in SECCO~1.
%In this context it is worth recalling that neither the Field~D Virgo dSphs or SECCO~1 can be %perceived in SDSS images. On the other hand the faint dwarfs in the SHIELD sample \citep{can11} are %unambiguously identified in SDSS images as small spots of diffuse light; none of them show a %morphology even barely similar to SECCO~1. For this reason we feel that the conclusion that SECCO~1 %is just another SHIELD galaxy, recently advanced by \citet{adams15}, is premature, at the present %stage of the analysis.

%------------------------FIG -----------------------------------
   \begin{figure*}
   \centering
   \includegraphics[width=\textwidth]{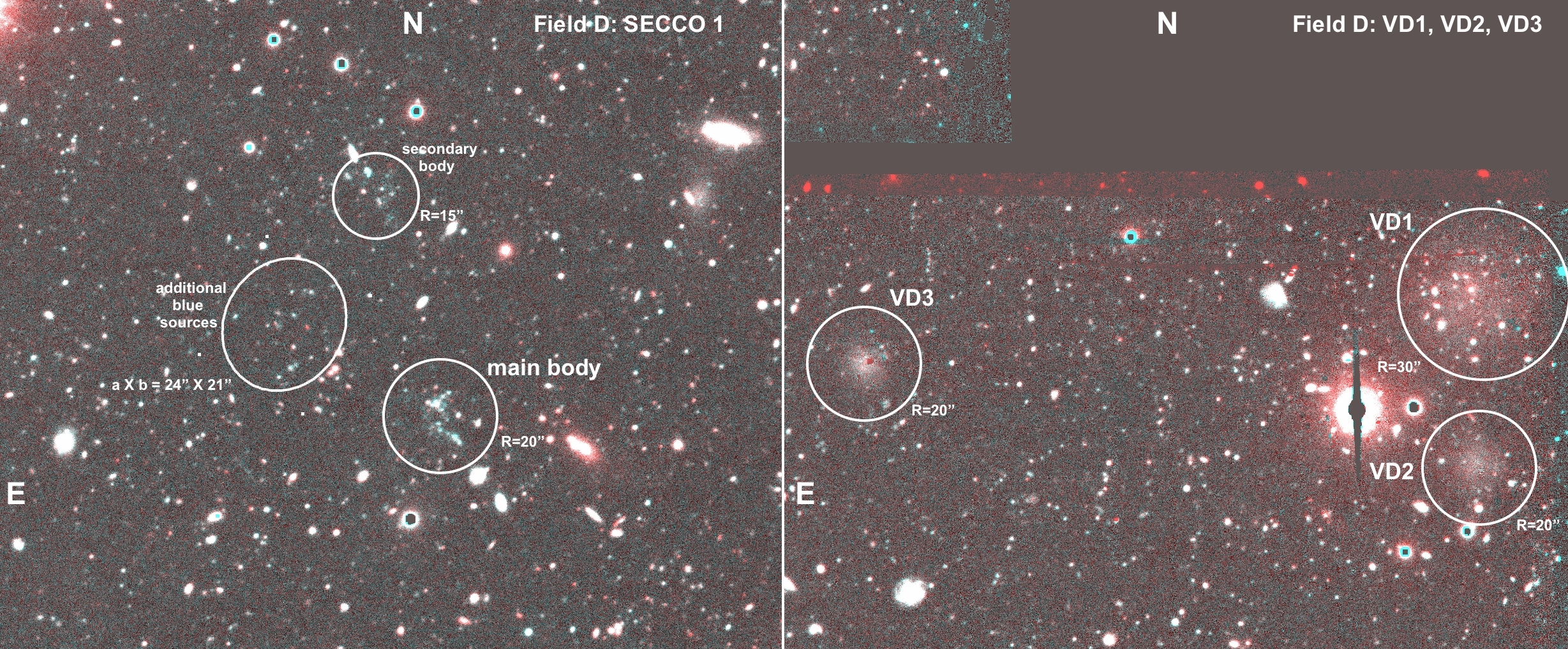}
     \caption{RGB color stacked images of SECCO Field~D centred on SECCO~1 (left panel) and VD1, VD2 and VD3 (right panel). The images have been obtained using the r-band image for the R channel and the g-band image for the G and B channels. The most remarkable structures have been labelled and enclosed with circles/ellipses whose size is reported.}
        \label{secco1_vd}
    \end{figure*}

%%%%%%%%%%%%%%%%%%%%%%%%%%%%%%%%%%%%%%%%%%%%%%%%%%%%%%%%%%%%%%%%%

\section{Summary and conclusions}
\label{conc}

We have complemented the search for stellar counterparts in 25 selected ALFALFA UCHVCs (A13), performed in Pap~I, with a wide set of experiments with synthetic dwarf galaxies, in order to quantitatively assess the sensitivity of the SECCO survey. We have explored a grid of models of old and metal-poor stellar systems in the range $-5.0\le M_V\le -10.0$. The stars of the synthetic dwarfs are distributed according to an exponential profile with three different $R_h$ values spanning the $\pm 1\sigma_{{\rm log}R_h}$ range about the mean $M_V$~vs.~$R_h$ relation as modelled by \citet[][based on local galaxies]{brasseur}. We analysed the sensitivity of stellar density maps in the range 0.25~Mpc$\le D\le$2.5~Mpc, and the sensitivity of visual inspection of the images in the range 0.25~Mpc$\le D\le$5.0~Mpc.

%------------------------FIG -----------------------------------
   \begin{figure}
   \centering
   \includegraphics[width=\columnwidth]{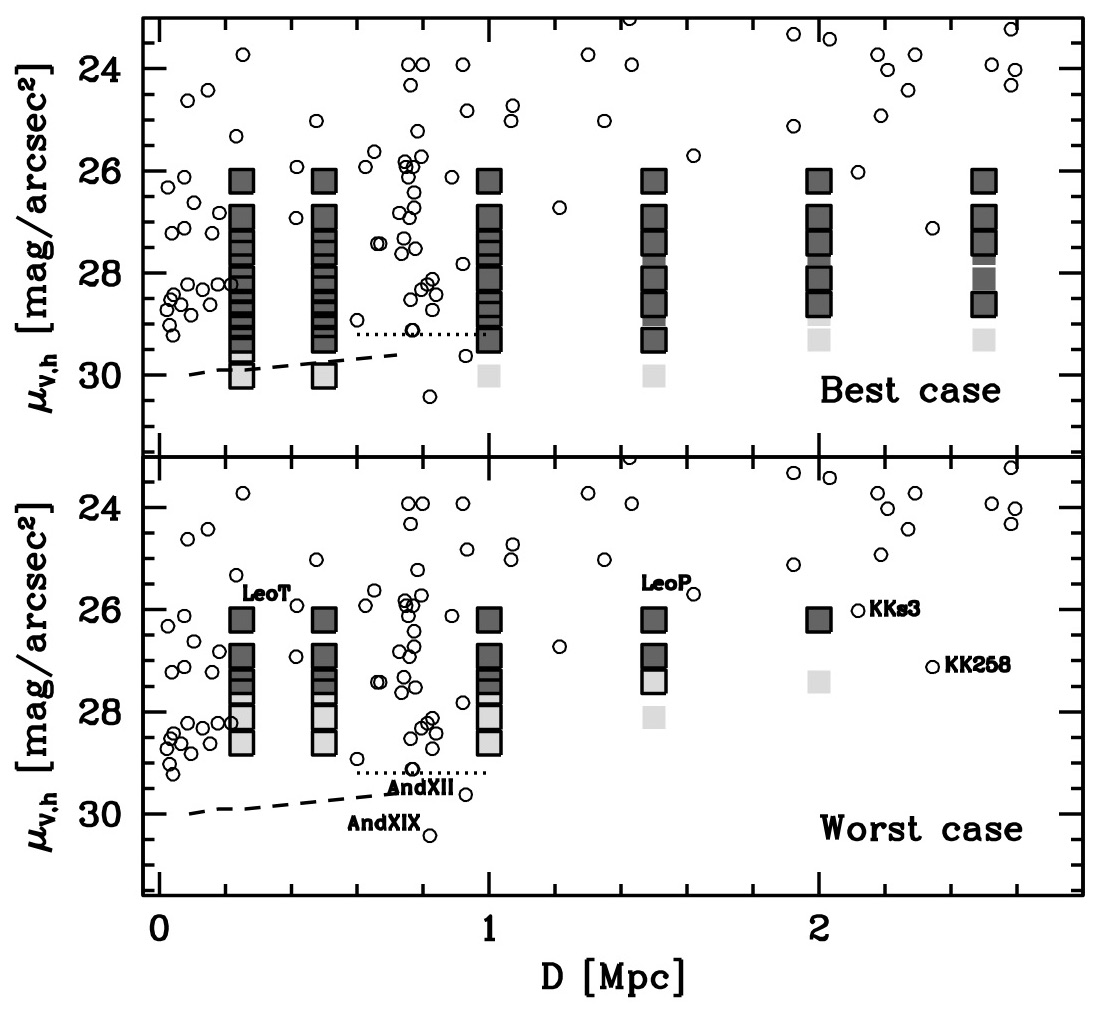}
     \caption{Sensitivity of density maps for {\em Best} (upper panel) and {\em Worst} (lower panel) cases in the Distance vs. surface brightness plane. The symbols are the same as in Fig.~\ref{sensB}. In particular, light open circles are known dwarf galaxies in the Local Volume, from \citet{mcc}. 
     {Central surface brightness values reported in the \citet{mcc} catalog have been transformed into $\mu_{V,h}$ assuming exponential profiles, an acceptable approximation in the considered range of luminosity \citep{mateo}}. We have labeled {\bf six} remarkable systems. The dashed line is the SDSS detection limit as a function of distance for resolved stellar systems from \citet[][their Tab.~3]{kopo8}; the dotted line is the detection limit of the PAndAS survey \citep{iba07} in the surroundings of M31, from \citet{brasseur}.}
        \label{sBmu}
    \end{figure}

%%%%%%%%%%%%%%%%%%%%%%%%%%%%%%%%%%%%%%%%%%%%%%%%%%%%%%%%%%%%%%%%%

We fully confirm and extend the results of Pap~I. In particular, we can now safely conclude that:

\begin{enumerate}

\item There is no dwarf galaxy with $R_h\le \bar R_h(M_v) +1\sigma_{{\rm log}R_h}$ and $M_V\le -10.0$, within $D\le 5.0$~Mpc, associated to any of the 25 A13 UCHVCs studied by SECCO.

\item There is no dwarf galaxy with $R_h\le \bar R_h(M_v) +1\sigma_{{\rm log}R_h}$ and $M_V\le -9.0$ within $D\le 1.0$~Mpc associated to any of the 25 A13 UCHVCs studied by SECCO.

\item If we consider only dwarfs having $R_h\le \bar R_h(M_v)$, there is no galaxy with $M_V\le -9.0$ within $D\le 2.0$~Mpc (as well as no galaxy with $M_V\le -8.0$ within $D\le 1.5$~Mpc) associated to any of the 25 A13 UCHVCs studied by SECCO.

\item If we consider the 9 SECCO Fields with the best data quality (namely, A, G, B, Y, R, K, U, C, L; see Pap~I)
we can conclude that there is no dwarf with $M_V\le -8.0$ within $D\le 2.5$~Mpc (as well as no galaxy with $M_V\le -7.0$ within $D\le 1.5$~Mpc) associated to the corresponding A13 UCHVCs.

\end{enumerate}

Thus we fully confirm the lack of $M_V\le -8.0$ local counterparts to ALFALFA UCHVCs already mentioned in Pap~I and, later, confirmed on the whole A13 sample but with 
shallower and less homogeneous data by \citep{sand15}.

In Fig.~\ref{sBmu} we provide a complementary view of the summary of our results presented in Figures~\ref{sensB} and \ref{sensI}, above, that may help to express our sensitivity limits in a more synthetic form, in the Distance vs. Surface Brightness plane. In {\em Best case} observations, we can detect as $\ge 5\sigma$ over-densities dwarfs with $\mu_{V,h}\le 30.0$~mag/arcsec$^2$ out to $D\le 1.5$~Mpc, and those with $\mu_{V,h}\le 29.5$~mag/arcsec$^2$ out to $D\le 2.5$~Mpc. In {\em Worst case} observations, we can detect as $\ge 5\sigma$ over-densities dwarfs with $\mu_{V,h}\le 28.8$~mag/arcsec$^2$ out to $D\le 1.0$~Mpc, and those with $\mu_{V,h}\le 27.6$~mag/arcsec$^2$ out to $D\le 2.5$~Mpc. More specifically, the comparison with the distribution of known dwarfs in the Local Volume \citep[from][]{mcc} shows that, for all the surveyed fields, independently of the quality of the available data:
\begin{itemize}

\item we would have detected any galaxy lying in range of distances and surface brightness inhabited by known LV dwarfs (except perhaps And~XII, And~XIX, KKs3, and KK258, with Worst case observations)

\item our sensitivity region extends into a range of low surface brightness and large distances (for 1.0~Mpc$\la D\la$~2.5~Mpc) that appears as largely unexplored, at present. 

\end{itemize}

In particular, Fig.~\ref{sBmu} shows that our survey is sensitive to dwarfs much fainter than the prototypical gas-rich faint dwarfs Leo~T and Leo~P.

%To put this conclusion into a simplified but more direct form, we can say that {\em there is no stellar system analogous to Leo~T or Leo~P associated with the ALFALFA UCHVCs targeted by SECCO}.
%If any of the putative UV counterparts found by \citet{donma} with a statistical procedure is indeed associated with the corresponding gas clouds they should be extremely faint and/or distant stellar systems, more similar to SECCO~1 than to normal star-forming galaxies. Indeed, in addition to SECCO~1, we found only two additional candidate stellar counterparts; both of them should be quite faint ($M_V\la -6$) or more distant than $D\sim$3.0~Mpc\footnote{One of them, over-density Q1, was presented in Pap~I. The other, over-density Y1, has been found as a $\ge 10\sigma$ over-density in density maps obtained considering only blue stars ($-0.6<(g-r)_0<=+0.2$) in SECCO catalogs, and will be described elsewhere.}. 

It is important to stress here that $M_V=-10.0$ corresponds to $8.5\times 10^5~L_{V,\sun}$, $M_V=-9.0$ to $3.4\times 10^5~L_{V,\sun}$, $M_V=-8.0$ to $1.3\times 10^5~L_{V,\sun}$, and $M_V=-7.0$ to $5\times 10^4~L_{V,\sun}$. Since star-forming galaxies have stellar mass-to-light ratios $M/L\le 1.0$, these translates into quite strong constraints in terms of total stellar mass of any stellar counterpart that went undetected in SECCO. For example, assuming a distance of 1.0~Mpc A13 concluded that ALFALFA UCHVCs span a range of \HI~ masses $10^5~M_{\sun}\la M_{\HI}\la 10^6~M_{\sun}$. This would imply that any associated stellar system not detected in SECCO would typically have $M_{\HI}/M_{*}\ga 10$, a value significantly higher than what found in normal star-forming dwarfs \citep[see, e.g.][]{can15,secco_l1,sand15} but quite typical of extremely metal-poor galaxies \citep[XMPs,][]{filho}. 

In conclusion, the results of our very deep and homogeneous survey concerning local counterparts of High Velocity Clouds is fully in line with the various unsuccessful attempts performed in the last two decades \citep[see, e.g.,][and references therein]{simon,will_count,hopp03,hopp07,siegel}. The novelty is that, on the other hand, some more distant counterpart begins to emerge \citep[Pap~I,][]{secco_l1}, especially among the GALFA-HI candidates \citep{sand15}. The follow up of these counterparts and further searches in better defined samples of candidates may provide crucial insight on the census and the evolution of very low mass star-forming dwarfs.

%%%%%%%%%%%%%%%%%%%%%%%%%%%%%%%%%%%%%%%%%%%%%%%%%%%%%%%%%%%%%%%%%
\begin{acknowledgements}

We are grateful to an anonymous referee for providing insightful comments and and precious suggestions for a clearer and more meaningful presentation of the results of this analysis.
We acknowledge the support from the LBT-Italian Coordination Facility for the execution of observations, data distribution and reduction.  

M.B. acknowledge the financial support from PRIN MIUR 2010-2011 project ``The
Chemical and Dynamical Evolution of the Milky Way and Local Group Galaxies'',
prot. 2010LY5N2T. 

G. Battaglia gratefully acknowledges support through a Marie-Curie action Intra European Fellowship, funded from the European Union Seventh Framework Program (FP7/2007-2013) under Grant agreement number PIEF-GA-2010-274151, as well as the financial support by the Spanish Ministry of Economy 
and Competitiveness (MINECO) under the Ram\'on y Cajal Program 
(RYC-2012-11537).

This research has made use of the SIMBAD database, operated at CDS, Strasbourg, France.
This research has made use of the NASA/IPAC Extragalactic Database (NED) which is operated by the Jet Propulsion Laboratory, California Institute of Technology, under contract with the National Aeronautics and Space Administration. 
This research has made use of NASA's Astrophysics Data System.

This research make use of SDSS data. Funding for the SDSS and SDSS- II has been provided by the Alfred P. Sloan Foundation, the Participating Institutions, the National Science Foundation, the US Department of Energy, the National Aeronautics and Space Administration, the Japanese Monbukagakusho, the Max Planck Society, and the Higher Education Funding Council for England. The SDSS Web Site is http:www.sdss.org. The SDSS is managed by the Astrophysical Research Consortium for the Participating Institutions. The Participating Institutions are the American Museum of Natural History, Astrophysical Institute Potsdam, University of Basel, University of Cambridge, Case Western Reserve University, University of Chicago, Drexel University, Fermilab, the Institute for Advanced Study, the Japan Participation Group, Johns Hopkins University, the Joint Institute for Nuclear Astrophysics, the Kavli Institute for Particle Astrophysics and Cosmology, the Korean Scientist Group, the Chinese Academy of Sciences (LAMOST), Los Alamos National Laboratory, the Max-Planck-Institute for Astronomy (MPIA), the Max-Planck- Institute for Astrophysics (MPA), New Mexico State University, Ohio State University, University of Pittsburgh, University of Portsmouth, Princeton University, the United States Naval Observatory, and the University of Washington.
\end{acknowledgements}

\bibliographystyle{apj}

\appendix

\section{Sensitivity of density maps: a summary table}
\label{app_dens}

In Table~\ref{classi} we summarise the results of our experiments for the detection synthetic dwarf galaxies with density maps described in Sect.~\ref{sensmap} and illustrated in Fig.~\ref{sensB} and Fig.~\ref{sensI}, above. 

\begin{table*}
  \begin{center}
  \caption{Classification of the over-density detections}
  \label{classi}
  \begin{tabular}{lcccccccc}
$M_V$ &  k$^a$ & D & $B_{10}$ & $B_{5}$ & $B_{CMD}$ & $I_{10}$ & $I_{5}$ & $I_{CMD}$\\
 \hline
 -9.0 & -1 & 0.25  &   1  &   1  &   1  &   1	&   1	&  1  \\
 -9.0 & -1 & 0.50  &   1  &   1  &   1  &   1	&   1	&  1  \\
 -9.0 & -1 & 1.00  &   1  &   1  &   1  &   1	&   1	&  1  \\
 -9.0 & -1 & 1.50  &   1  &   1  &   1  &   1	&   1	&  1  \\
 -9.0 & -1 & 2.00  &   1  &   1  &   1  &   1	&   1	&  1  \\
 -9.0 & -1 & 2.50  &   1  &   1  &   1  &   0	&   0	&  0  \\
 -9.0 &  0 & 0.25  &   1  &   1  &   1  &   1	&   1	&  1  \\
 -9.0 &  0 & 0.50  &   1  &   1  &   1  &   1	&   1	&  1  \\
 -9.0 &  0 & 1.00  &   1  &   1  &   1  &   1	&   1	&  1  \\
 -9.0 &  0 & 1.50  &   1  &   1  &   1  &   0	&   1	&  1  \\
 -9.0 &  0 & 2.00  &   1  &   1  &   1  &   0	&   1	&  0  \\
 -9.0 &  0 & 2.50  &   1  &   1  &   1  &   0	&   0	&  0  \\
 -9.0 & +1 & 0.25  &   1  &   1  &   1  &   0	&   1	&  1  \\
 -9.0 & +1 & 0.50  &   1  &   1  &   1  &   0	&   1	&  1  \\
 -9.0 & +1 & 1.00  &   1  &   1  &   1  &   0	&   1	&  1  \\
 -9.0 & +1 & 1.50  &   1  &   1  &   1  &   0	&   0	&  0  \\
 -9.0 & +1 & 2.00  &   1  &   1  &   1  &   0	&   0	&  0  \\
 -9.0 & +1 & 2.50  &   1  &   1  &   1  &   0	&   0	&  0  \\
 -8.0 & -1 & 0.25  &   1  &   1  &   1  &   1	&   1	&  1  \\
 -8.0 & -1 & 0.50  &   1  &   1  &   1  &   1	&   1	&  1  \\
 -8.0 & -1 & 1.00  &   1  &   1  &   1  &   1	&   1	&  1  \\
 -8.0 & -1 & 1.50  &   1  &   1  &   1  &   1	&   1	&  1  \\
 -8.0 & -1 & 2.00  &   1  &   1  &   1  &   0	&   0	&  0  \\
 -8.0 & -1 & 2.50  &   1  &   1  &   1  &   0	&   0	&  0  \\
 -8.0 &  0 & 0.25  &   1  &   1  &   1  &   0	&   1	&  1  \\
 -8.0 &  0 & 0.50  &   1  &   1  &   1  &   0	&   1	&  1  \\
 -8.0 &  0 & 1.00  &   1  &   1  &   1  &   0	&   1	&  1  \\
 -8.0 &  0 & 1.50  &   1  &   1  &   1  &   0	&   1	&  0  \\
 -8.0 &  0 & 2.00  &   1  &   1  &   1  &   0	&   0	&  0  \\
 -8.0 &  0 & 2.50  &   1  &   1  &   0  &   0	&   0	&  0  \\
 -8.0 & +1 & 0.25  &   1  &   1  &   1  &   0	&   0	&  0  \\
 -8.0 & +1 & 0.50  &   1  &   1  &   1  &   0	&   0	&  0  \\
 -8.0 & +1 & 1.00  &   1  &   1  &   1  &   0	&   0	&  0  \\
 -8.0 & +1 & 1.50  &   1  &   1  &   1  &   0	&   0	&  0  \\
 -8.0 & +1 & 2.00  &   0  &   1  &   0  &   0	&   0	&  0  \\
 -8.0 & +1 & 2.50  &   0  &   1  &   0  &   0	&   0	&  0  \\
 -7.0 & -1 & 0.25  &   1  &   1  &   1  &   0	&   1	&  1  \\
 -7.0 & -1 & 0.50  &   1  &   1  &   1  &   0	&   1	&  1  \\
 -7.0 & -1 & 1.00  &   1  &   1  &   1  &   0	&   1	&  1  \\
 -7.0 & -1 & 1.50  &   1  &   1  &   0  &   0	&   0	&  0  \\
 -7.0 & -1 & 2.00  &   1  &   1  &   0  &   0	&   0	&  0  \\
 -7.0 & -1 & 2.50  &   1  &   1  &   0  &   0	&   0	&  0  \\
 -7.0 &  0 & 0.25  &   1  &   1  &   1  &   0	&   0	&  0  \\   
 -7.0 &  0 & 0.50  &   1  &   1  &   1  &   0	&   0	&  0  \\
 -7.0 &  0 & 1.00  &   1  &   1  &   1  &   0	&   0	&  0  \\
 -7.0 &  0 & 1.50  &   1  &   1  &   0  &   0	&   0	&  0  \\
 -7.0 &  0 & 2.00  &   0  &   1  &   0  &   0	&   0	&  0  \\
 -7.0 &  0 & 2.50  &   0  &   0  &   0  &   0	&   0	&  0  \\
 -7.0 & +1 & 0.25  &   0  &   1  &   1  &   0	&   0	&  0  \\
 -7.0 & +1 & 0.50  &   0  &   1  &   1  &   0	&   0	&  0  \\
 -7.0 & +1 & 1.00  &   0  &   1  &   0  &   0	&   0	&  0  \\
 -7.0 & +1 & 1.50  &   0  &   1  &   0  &   0	&   0	&  0  \\
 -7.0 & +1 & 2.00  &   0  &   0  &   0  &   0	&   0	&  0  \\
 -7.0 & +1 & 2.50  &   0  &   0  &   0  &   0	&   0	&  0  \\
 ... continue&&&&&&&&\\
\hline
\end{tabular} 
%\tablefoot{$^a$ indicates compact ($k=-1$), average ($k=0$) or diffuse ($k=+1$) models,
%where $R_h=\bar R_h(M_v) +k\sigma_{{\rm log}R_h}$.\\
%$B_{10}(I_{10})=1$ indicates a $\ge 10\sigma$ detection in Field~B(Field~I).\\
%$B_{5}(I_{5})=1$ indicates a $\ge 5\sigma$ and $< 10\sigma$ detection in Field~B(Field~I).\\
%$B_{CMD}(I_{CMD}=1$ indicates that the over-density can be recognised as a real galaxy by inspection of 
%the CMD.}
\end{center}
\end{table*}
%%%%%%%%%%%%%%%%%%%%%%%%%%%%%%%%%%%%%%%%%%%%%%%%%%%%%%%%%%%%%%%%%%%%%%%%%%%%%%%%%%%%%%%%%%%%%%%%%%%%%%%%%%%%

%%%%%%%%%%%%%%%%%%%%%%%%%%%%%%%%%%%%%%%%%%%%%%%%%%%%%%%%%%%%%%%%%%%%%%%%%%%%%%%%%%%%%%%%%%%%%%%%%%%%%%%%%%%%
\setcounter{table}{0}
\begin{table*}
  \begin{center}
  \caption{Classification of the over-density detections}
  \label{classi}
  \begin{tabular}{lcccccccc}
$M_V$ &  k$^a$ & D & $B_{10}$ & $B_{5}$ & $B_{CMD}$ & $I_{10}$ & $I_{5}$ & $I_{CMD}$\\
 \hline
 -6.0 & -1 & 0.25  &   1  &   1  &   1  &   0	&   0	&  0  \\
 -6.0 & -1 & 0.50  &   0  &   1  &   1  &   0	&   0	&  0  \\
 -6.0 & -1 & 1.00  &   0  &   1  &   0  &   0	&   0	&  0  \\
 -6.0 & -1 & 1.50  &   0  &   1  &   0  &   0	&   0	&  0  \\
 -6.0 & -1 & 2.00  &   0  &   0  &   0  &   0	&   0	&  0  \\
 -6.0 & -1 & 2.50  &   0  &   0  &   0  &   0	&   0	&  0  \\
 -6.0 &  0 & 0.25  &   0  &   1  &   1  &   0	&   0	&  0  \\
 -6.0 &  0 & 0.50  &   0  &   0  &   0  &   0	&   0	&  0  \\
 -6.0 &  0 & 1.00  &   0  &   0  &   0  &   0	&   0	&  0  \\
 -6.0 &  0 & 1.50  &   0  &   0  &   0  &   0	&   0	&  0  \\
 -6.0 &  0 & 2.00  &   0  &   0  &   0  &   0	&   0	&  0  \\
 -6.0 &  0 & 2.50  &   0  &   0  &   0  &   0	&   0	&  0  \\
 -6.0 & +1 & 0.25  &   0  &   0  &   0  &   0	&   0	&  0  \\
 -6.0 & +1 & 0.50  &   0  &   0  &   0  &   0	&   0	&  0  \\
 -6.0 & +1 & 1.00  &   0  &   0  &   0  &   0	&   0	&  0  \\
 -6.0 & +1 & 1.50  &   0  &   0  &   0  &   0	&   0	&  0  \\
 -6.0 & +1 & 2.00  &   0  &   0  &   0  &   0	&   0	&  0  \\
 -6.0 & +1 & 2.50  &   0  &   0  &   0  &   0	&   0	&  0  \\
 -5.0 & -1 & 0.25  &   0  &   1  &   1  &   0	&   0	&  0  \\
 -5.0 & -1 & 0.50  &   0  &   1  &   1  &   0	&   0	&  0  \\
 -5.0 & -1 & 1.00  &   0  &   1  &   0  &   0	&   0	&  0  \\
 -5.0 & -1 & 1.50  &   0  &   0  &   0  &   0	&   0	&  0  \\
 -5.0 & -1 & 2.00  &   0  &   0  &   0  &   0	&   0	&  0  \\
 -5.0 & -1 & 2.50  &   0  &   0  &   0  &   0	&   0	&  0  \\

\hline
\end{tabular} 
\tablefoot{$^a$ indicates compact ($k=-1$), average ($k=0$) or diffuse ($k=+1$) models,
where $R_h=\bar R_h(M_v) +k\sigma_{{\rm log}R_h}$.\\
$B_{10}(I_{10})=1$ indicates a $\ge 10\sigma$ detection in Best(Worst) case observations (Field~B and Field~I, respectively).\\
$B_{5}(I_{5})=1$ indicates a $\ge 5\sigma$ and $< 10\sigma$ detection in Best(Worst) case observations (Field~B and Field~I, respectively).\\
$B_{CMD}(I_{CMD})=1$ indicates that the inspection of the CMD provide significant support to the classification of the detected over-density as a dwarf galaxy.}
\end{center}
\end{table*}
%%%%%%%%%%%%%%%%%%%%%%%%%%%%%%%%%%%%%%%%%%%%%%%%%%%%%%%%%%%%%%%%%%%%%%%%%%%%%%%%%%%%%%%%%%%%%%%%%%%%%%%%%%%%
\clearpage

\section{A small Atlas of synthetic dwarfs}
\label{app_ima}

In this appendix we provide images of synthetic galaxies exploring the most relevant cases for establishing the sensitivity limits of our survey. We feel that this small atlas of synthetic dwarfs can be of general utility to interpret images of candidate dwarfs obtained with 8m class telescopes. For a given galaxy model (identified by its $M_V$ and $R_h$ values), and for a given observational condition (Field~B: best conditions, Field~I: worst conditions) we simulate the galaxy image for five different distances (D=0.5, 1.0, 2.0, 3.0, and 5.0~Mpc).
This is represented, in the following, with six-panels figures with the stamp images of the given model for the five considered distances and a sixth panel showing the original empty image. In each of the six panels, the characteristics of the model as well as the considered value of the distance are clearly reported. A circle with radius equal to $R_h$ is superimposed and its value in arcsec is reported. Usually the images in the last four stamp-size panels have the same scale, while the first two are zoomed out to include the radius=$R_h$ circle.  
The lower set of panels in Fig.~\ref{imaI10} shows that a $M_V=-10.0$ would have been visually detected in SECCO images even in fields with the worst image quality and for {\em diffuse} models, i.e. dwarfs with $R_h=\bar R_h(M_v) +1\sigma_{{\rm log}R_h}$.

%%%%%%%%%%%%%%%%%%%%%%%%%%%%%%%%%%%%%%%%%%%%%%%%%%%%%%%%%%%%%%%%%
%------------------------FIG 3-----------------------------------
   \begin{figure*}
   \centering
   \includegraphics[width=\textwidth]{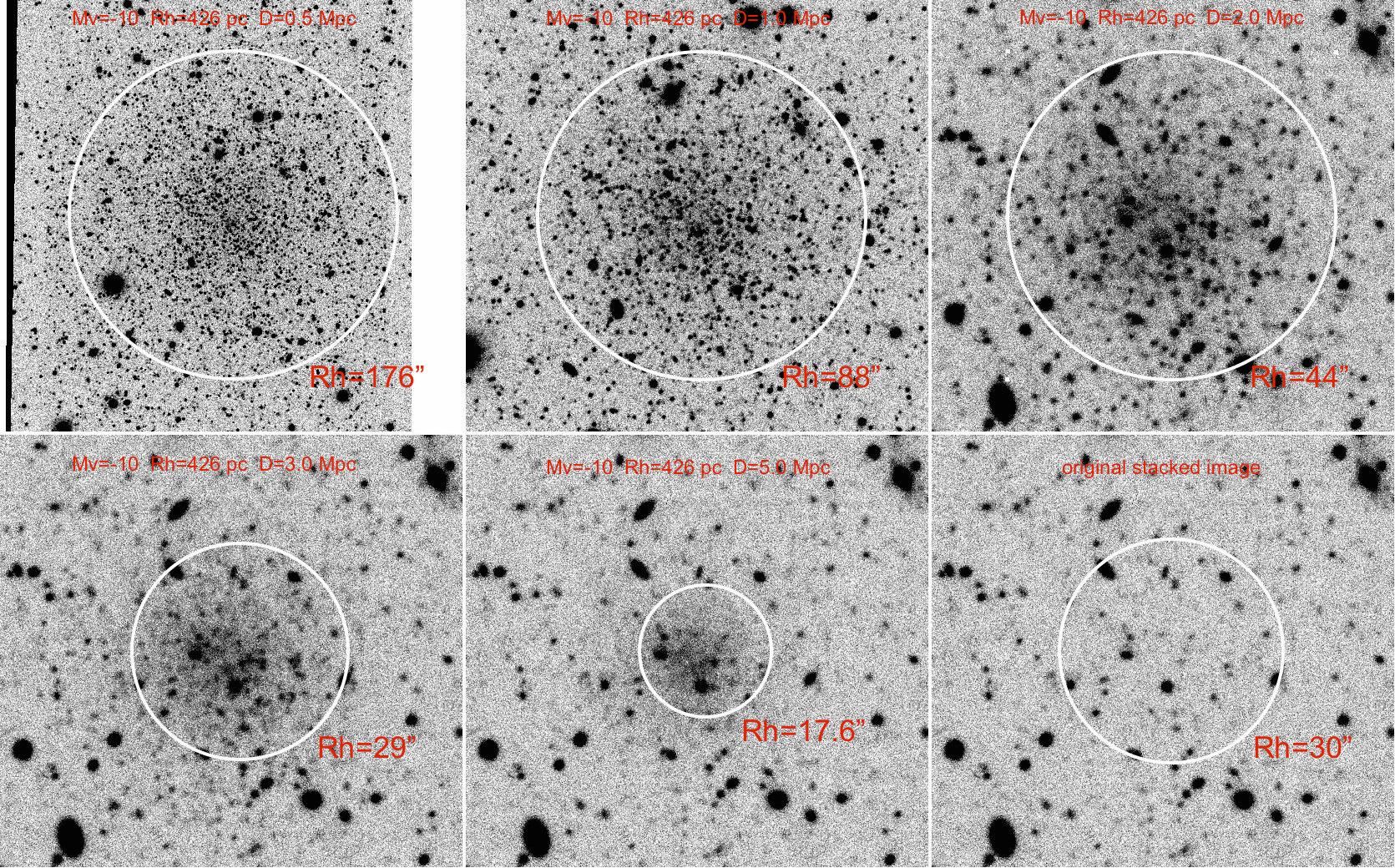}
   \includegraphics[width=\textwidth]{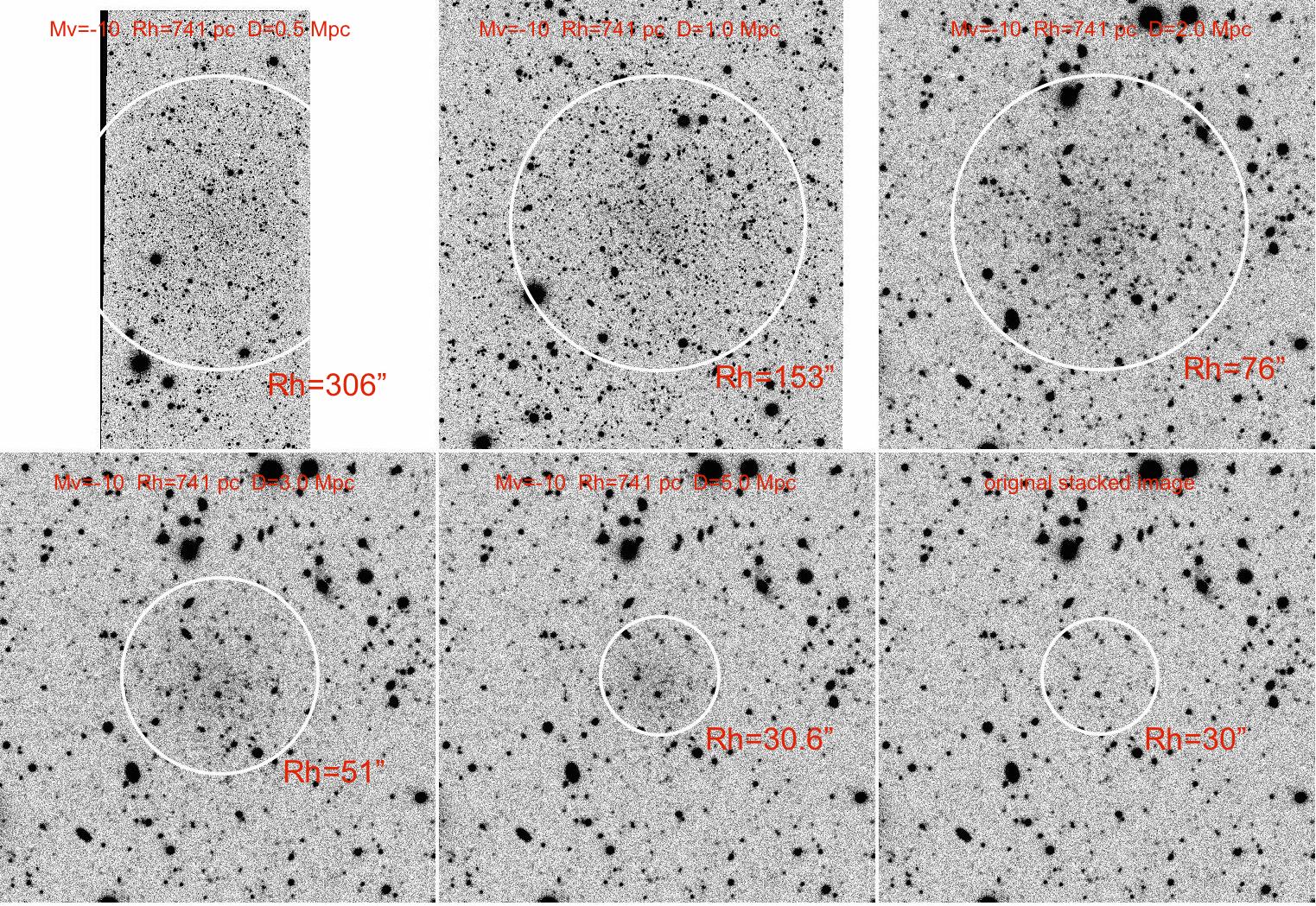}
     \caption{Worst case observations, $M_V=-10.0$, average (upper series of panels) and {\em diffuse} models (lower series of panels).}
        \label{imaI10}
    \end{figure*}

%\clearpage
%%%%%%%%%%%%%%%%%%%%%%%%%%%%%%%%%%%%%%%%%%%%%%%%%%%%%%%%%%%%%%%%%

%%%%%%%%%%%%%%%%%%%%%%%%%%%%%%%%%%%%%%%%%%%%%%%%%%%%%%%%%%%%%%%%%
%------------------------FIG 3-----------------------------------
   \begin{figure*}
   \centering
   \includegraphics[width=\textwidth]{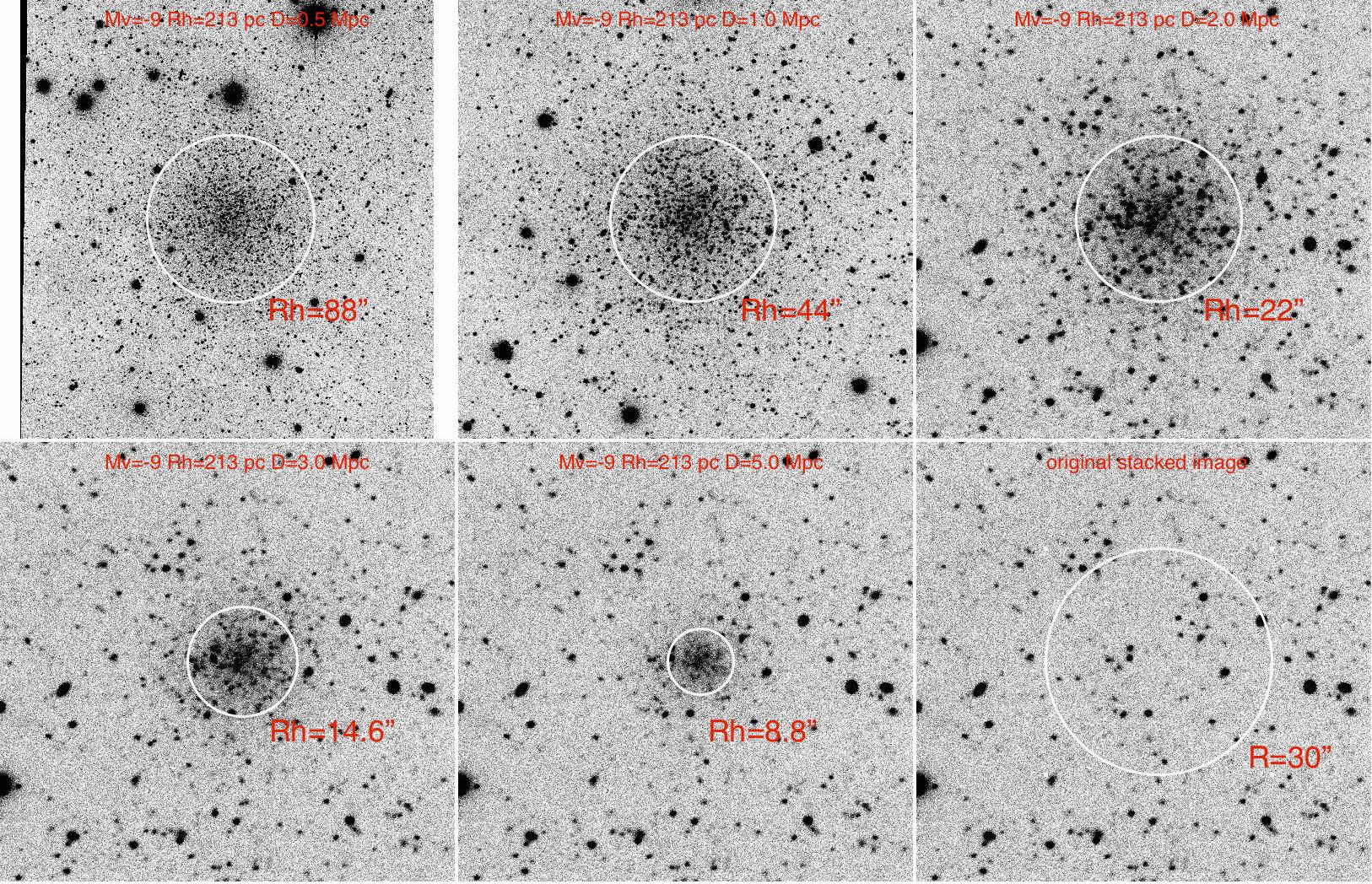}
   \includegraphics[width=\textwidth]{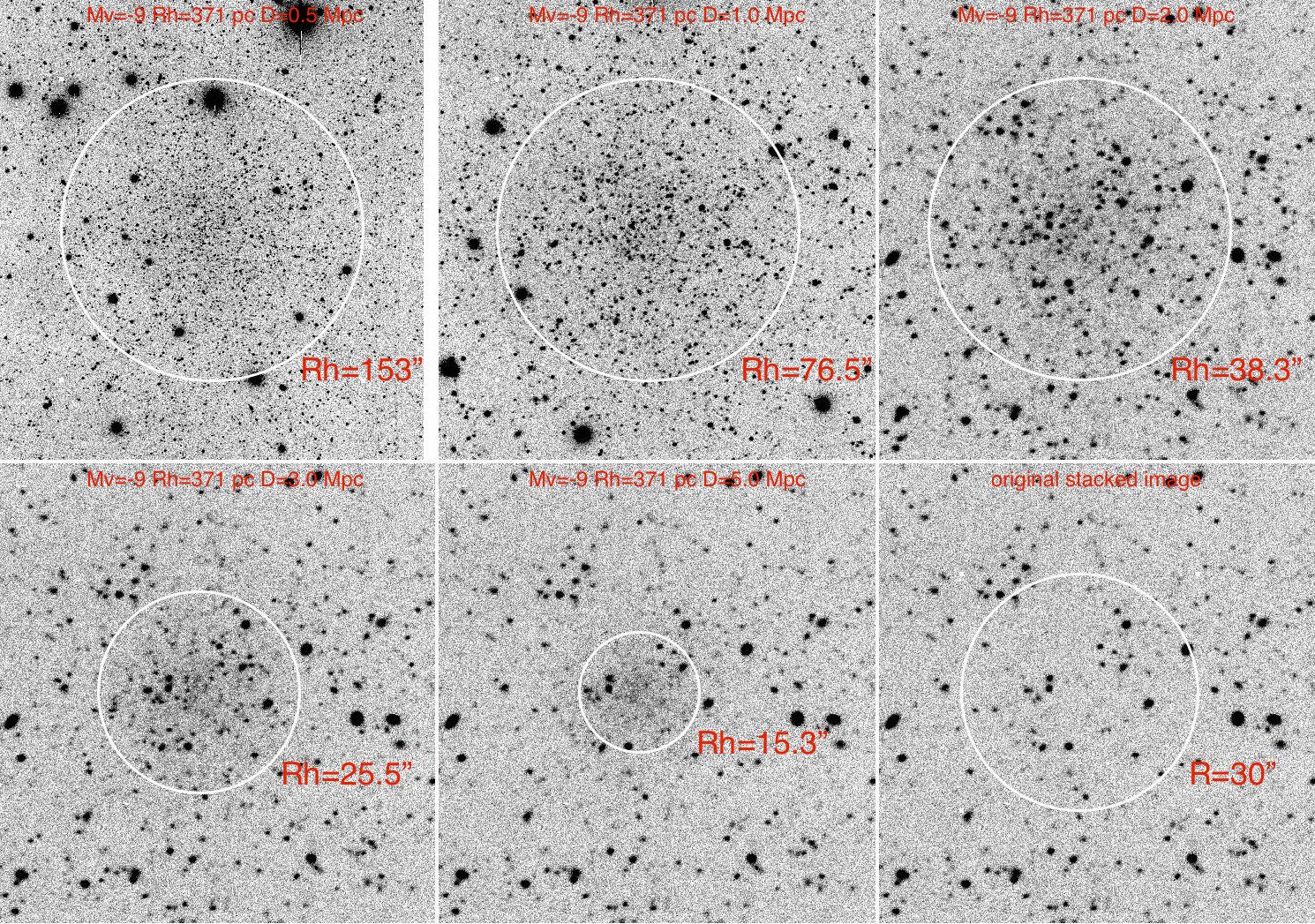}
     \caption{Best case observations, $M_V=-9.0$, compact (upper series of panels)  and average models (lower series of panels).}
        \label{imaB9}
    \end{figure*}

%\clearpage
%%%%%%%%%%%%%%%%%%%%%%%%%%%%%%%%%%%%%%%%%%%%%%%%%%%%%%%%%%%%%%%%%
%%%%%%%%%%%%%%%%%%%%%%%%%%%%%%%%%%%%%%%%%%%%%%%%%%%%%%%%%%%%%%%%%
%------------------------FIG 3-----------------------------------
   \begin{figure*}
   \centering
   \includegraphics[width=\textwidth]{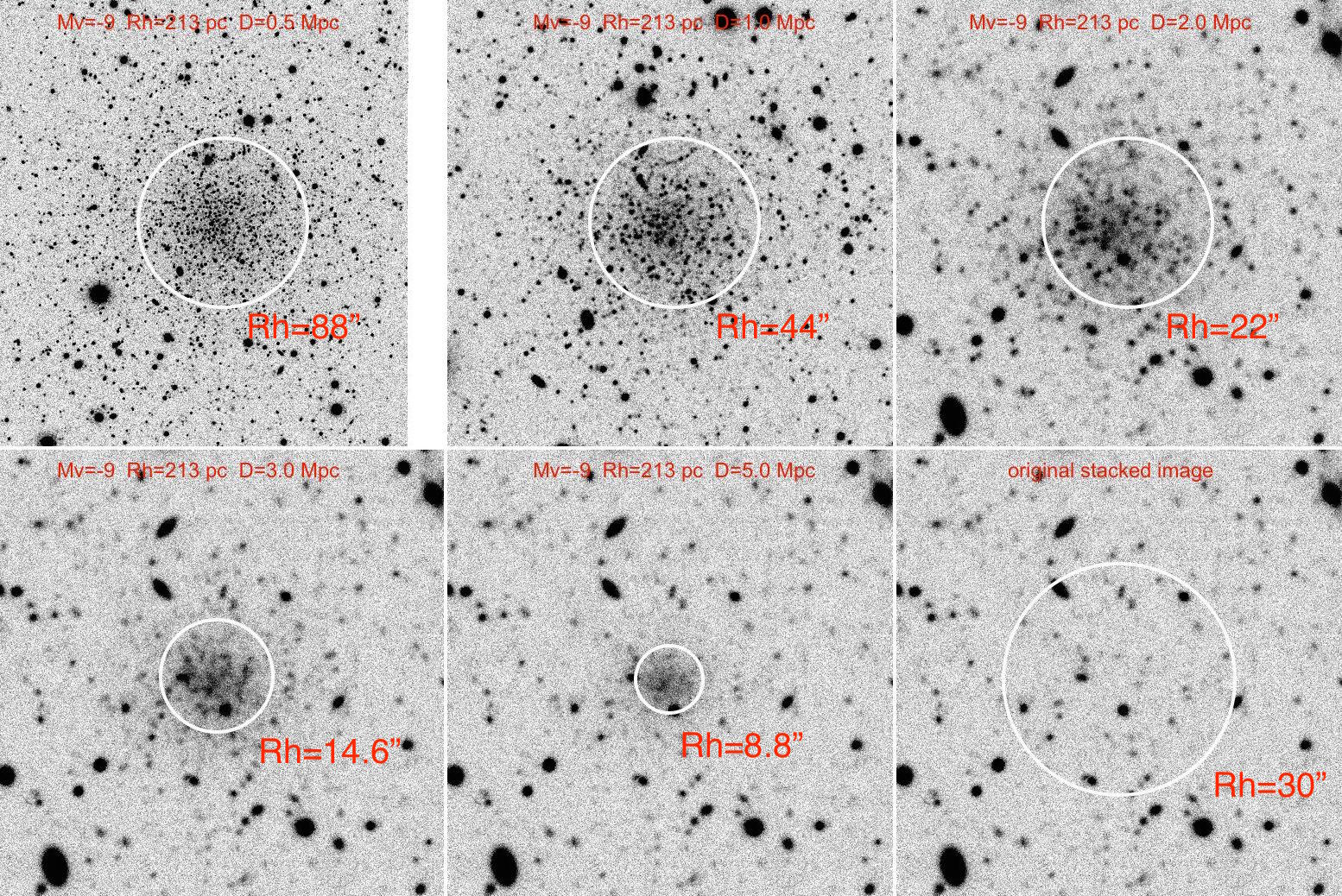}
   \includegraphics[width=\textwidth]{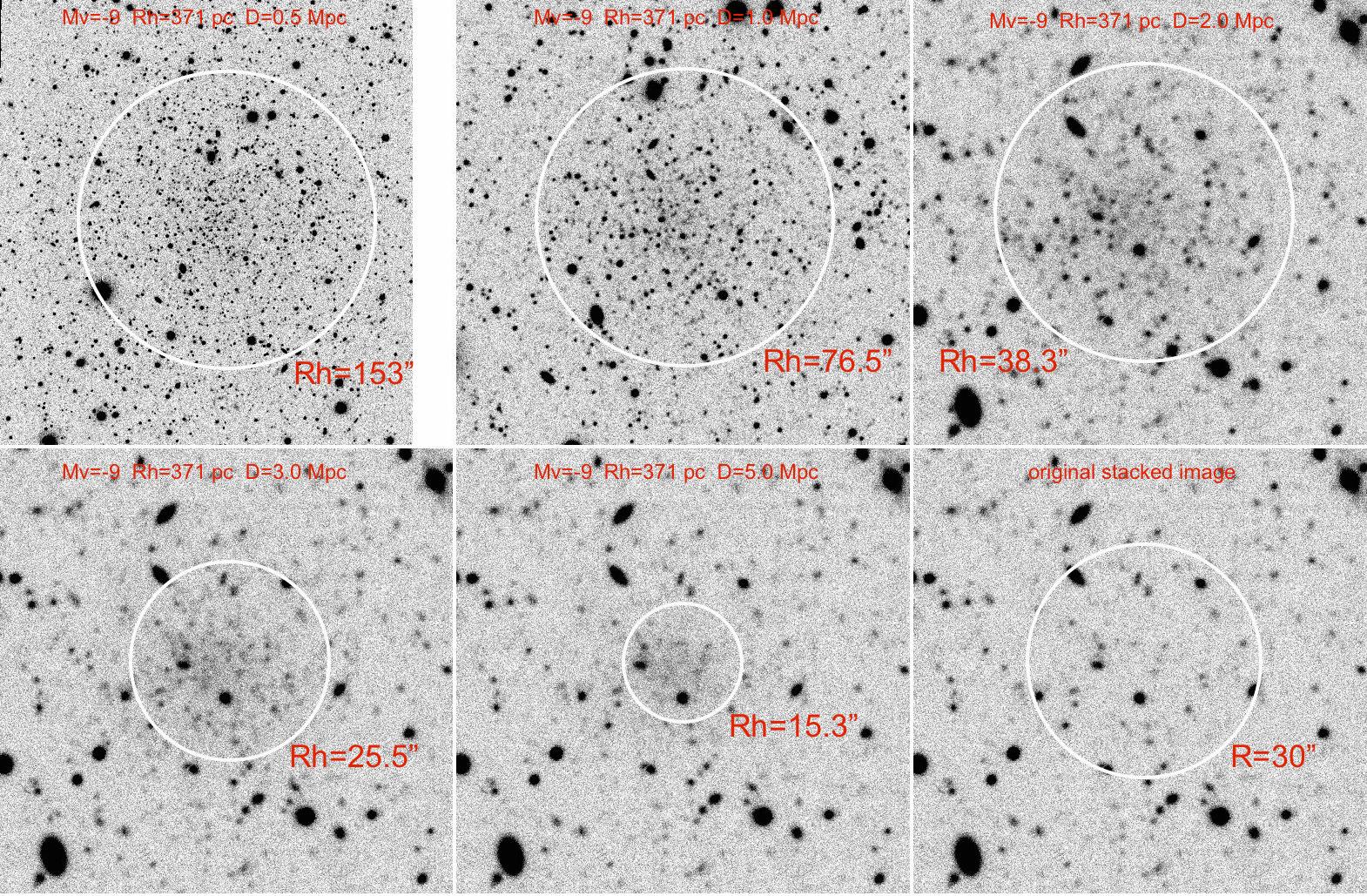}
     \caption{Worst case observations, $M_V=-9.0$, compact (upper series of panels)  and average models (lower series of panels). }
        \label{imaI9}
    \end{figure*}

%\clearpage
%%%%%%%%%%%%%%%%%%%%%%%%%%%%%%%%%%%%%%%%%%%%%%%%%%%%%%%%%%%%%%%%%

%%%%%%%%%%%%%%%%%%%%%%%%%%%%%%%%%%%%%%%%%%%%%%%%%%%%%%%%%%%%%%%%%
%------------------------FIG 3-----------------------------------
   \begin{figure*}
   \centering
   \includegraphics[width=\textwidth]{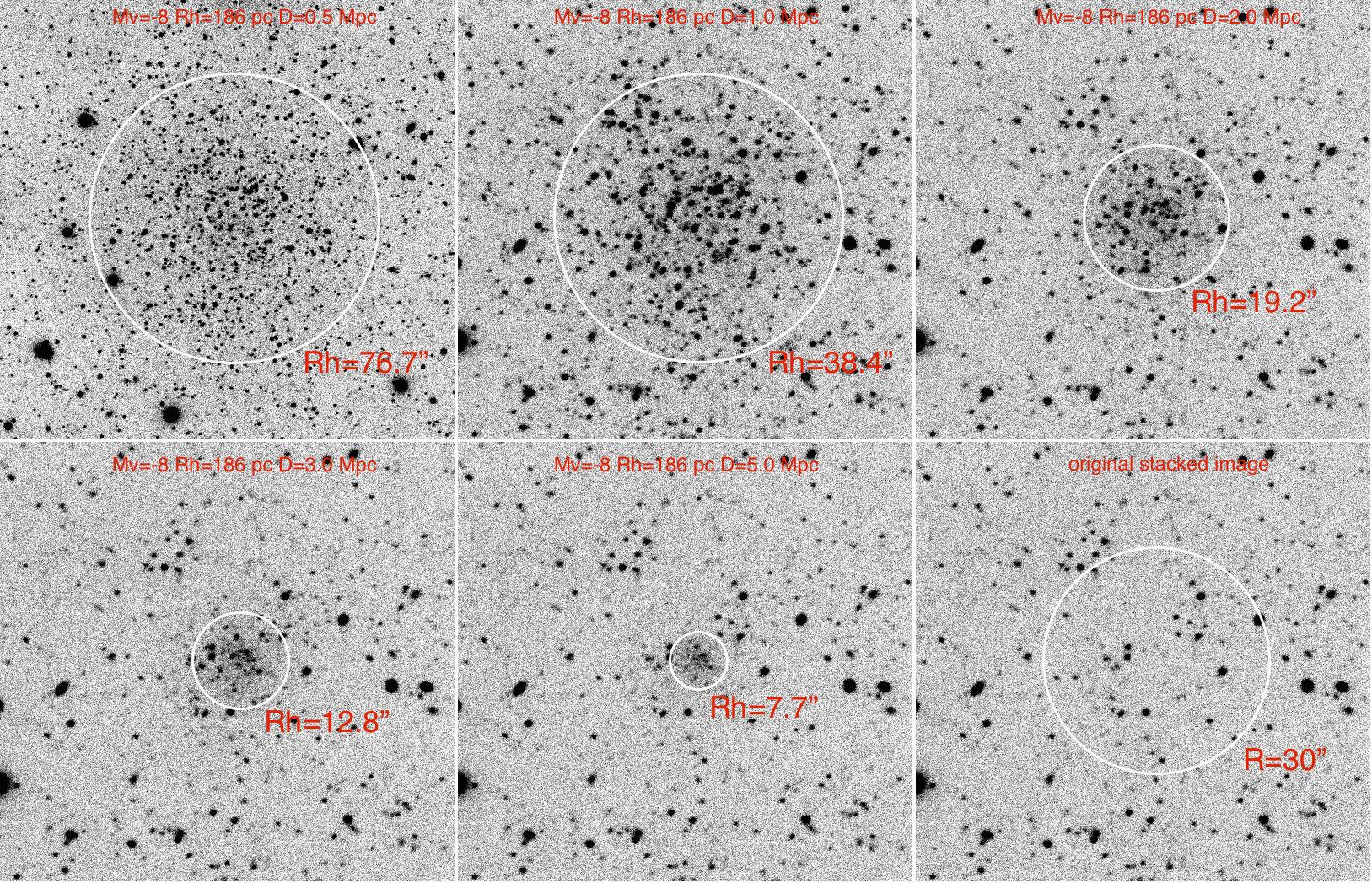}
   \includegraphics[width=\textwidth]{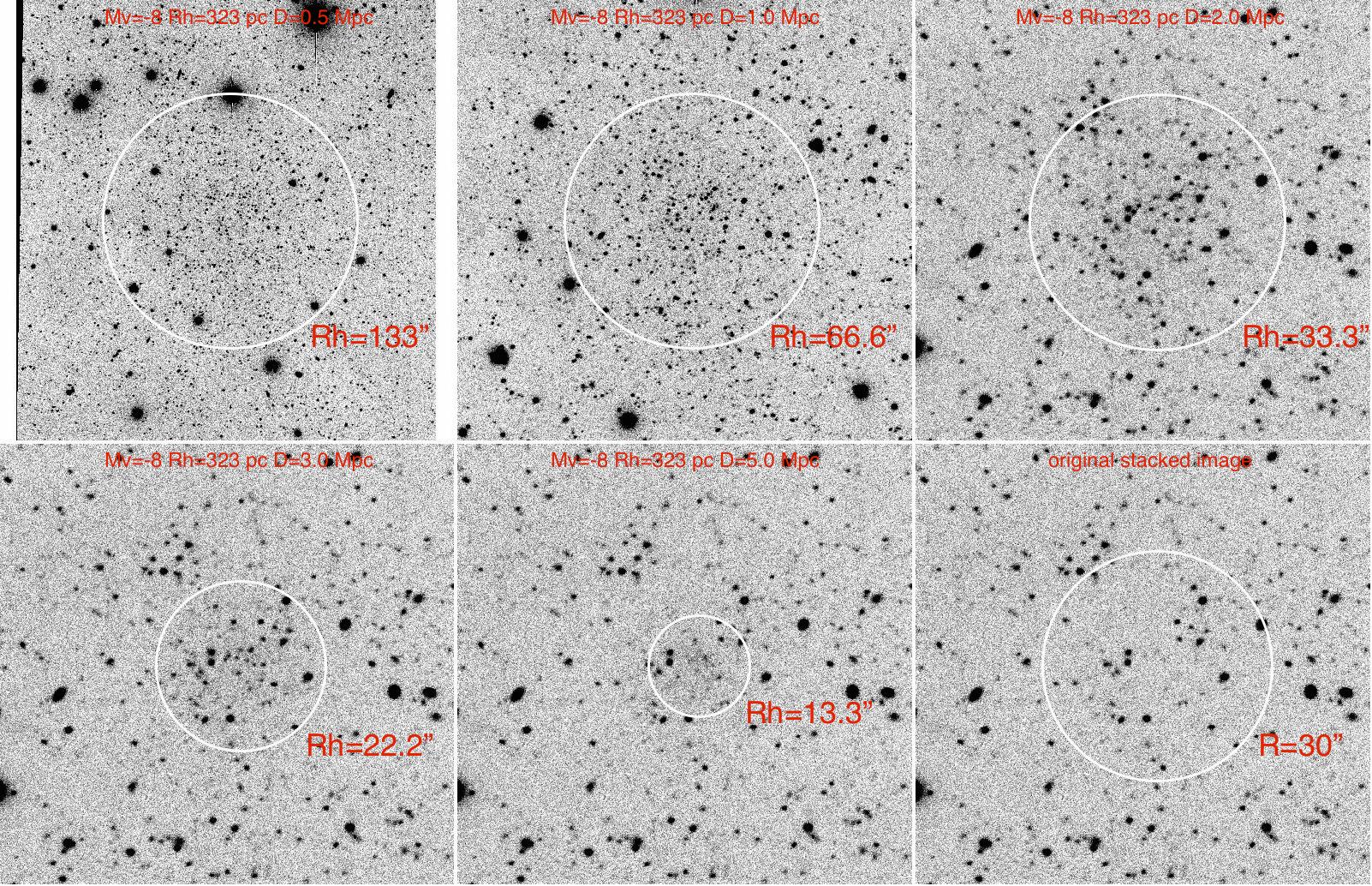}
     \caption{Best case observations, $M_V=-8.0$, compact (upper series of panels)  and average models (lower series of panels). }
        \label{imaB8}
    \end{figure*}

%\clearpage
%%%%%%%%%%%%%%%%%%%%%%%%%%%%%%%%%%%%%%%%%%%%%%%%%%%%%%%%%%%%%%%%%

%%%%%%%%%%%%%%%%%%%%%%%%%%%%%%%%%%%%%%%%%%%%%%%%%%%%%%%%%%%%%%%%%
%------------------------FIG 3-----------------------------------
   \begin{figure*}
   \centering
   \includegraphics[width=\textwidth]{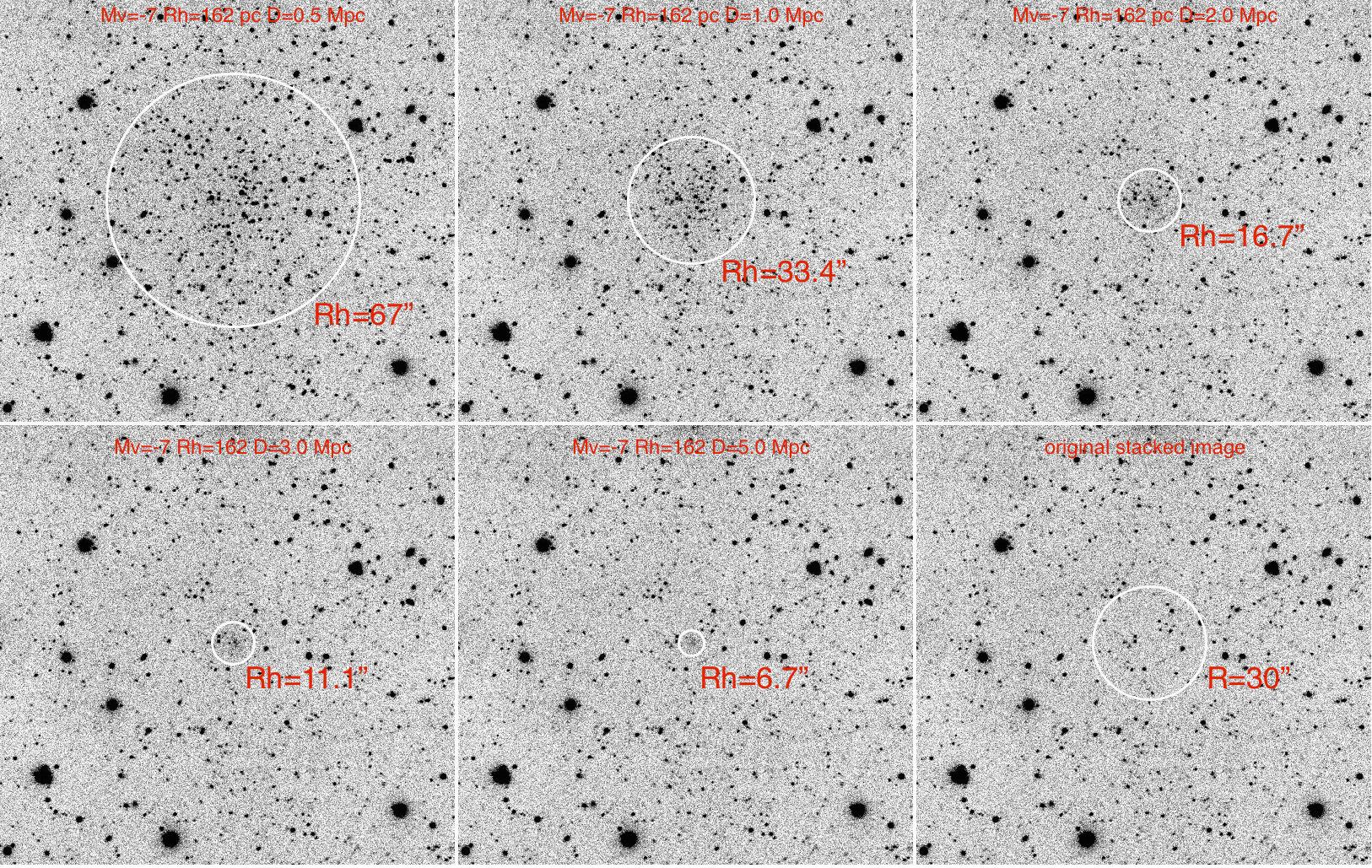}
   \includegraphics[width=\textwidth]{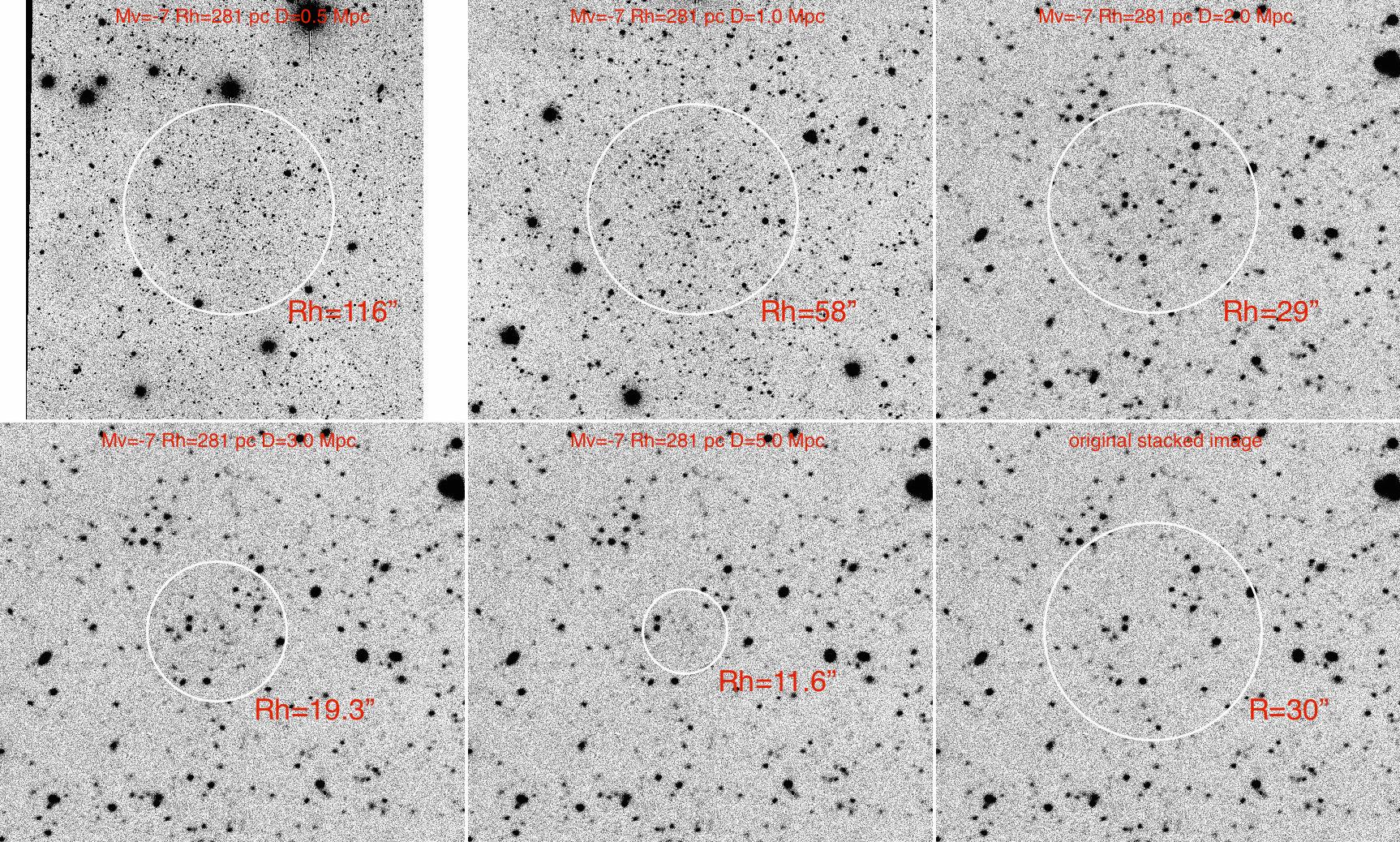}
     \caption{Best case observations, $M_V=-7.0$, compact (upper series of panels)  and average models (lower series of panels). }
        \label{imaB7}
    \end{figure*}

%\clearpage
%%%%%%%%%%%%%%%%%%%%%%%%%%%%%%%%%%%%%%%%%%%%%%%%%%%%%%%%%%%%%%%%%

%\newpage

%%%%%%%%%%%%%%%%%%%%%%%%%%%%%%%%%%%%%%%%%%%%%%%%%%%%%%%%%%%%%%%%%%%%%%%%%%%%%%%%%%%%%%%%%%%%%%%%%%%%%%%%%%%%

%%%%%%%%%%%%%%%%%%%%%%%%%%%%%%%%%%%%%%%%%%%%%%%%%%%%%%%%%%%%%%%%%%%%%%%%%%%%%%%%%%%%%%%%%%%%%%%%%%%%%%%%%%%%

\end{document}